\documentclass[letterpaper,preprint]{article}
\pdfoutput=1
\usepackage{jheppub}
\usepackage{color,latexsym, array,multirow,  verbatim}
\usepackage[config, singlelinecheck=true]{caption,subfig}
\usepackage{url}
\newcommand{\pair}[2]{\langle #1  #2 \rangle}
\newcommand{\tev}{~\text{TeV}}
\newcommand{\gev}{~\text{GeV}}

\newcommand{\ord}[1]{\mathcal{O}\left(#1 \right)} %of the order of

\newcommand{\dmin}{d^{\text{min}}}
\newcommand{\kt}{k_T}
\newcommand{\CA}{\text{C/A}}

\title{On Statistical Aspects of Qjets}

\author[a]{Stephen D.~Ellis,}
\author[b]{Andrew Hornig,}
\author[c]{David Krohn,}
\author[b,d]{and Tuhin S.~Roy,}

\affiliation[a]{Department of Physics, University of Washington, Seattle WA, 98195, USA}
\affiliation[b]{Theoretical Division T-2,
        Los Alamos National Laboratory,
	Los Alamos, NM,  87544, USA}
\affiliation[c]{Department of Physics, Harvard University, Cambridge MA, 02138, USA}
\affiliation[d]{Department of Theoretical Physics, Tata Institute of Fundamental Research, Mumbai 400005, India}

\date{\today}
\abstract{
The process by which jet algorithms construct jets and subjets is inherently ambiguous and equally well motivated algorithms often return very different answers.  The Qjets procedure was introduced by the authors to account for this ambiguity by considering many reconstructions of a jet at once, allowing one to assign a weight to each interpretation of the jet. Employing these weighted interpretations leads to an improvement in the statistical stability of many measurements.  Here we explore in detail the statistical properties of these sets of weighted measurements and demonstrate how they can be used to improve the reach of  jet-based studies.
}
\preprint{LA-UR-14-27250, TIFR/TH/14-23 }

\begin{document}

\maketitle

\flushbottom

%------------------------------------------------------------------------------------------------------------------
%\linenumbers
\section{\label{sec:intro}Introduction}

Jets arise in many important scattering processes encountered at the LHC.  Therefore, much experimental and theoretical effort has  recently gone into creating better tools for handling them. Techniques now exist to identify jets arising from the decay of boosted heavy particles~\cite{Seymour:1993mx, Butterworth:2007ke, Brooijmans:1077731,Butterworth:2008iy, Butterworth:2009qa, Kaplan:2008ie,Thaler:2008ju, Almeida:2008yp, Plehn:2009rk, Thaler:2010tr, Soper:2011cr, Englert:2011iz, Thaler:2011gf, Plehn:2011tg, Jankowiak:2011qa, Ellis:2012sd, Ellis:2012zp, Soper:2012pb, Larkoski:2013eya, Larkoski:2014gra, Ortiz:2014iza}, to remove unwanted radiation from jets~\cite{Butterworth:2008iy, Butterworth:2008sd, Butterworth:2008tr, Ellis:2009su, Ellis:2009me, Krohn:2009th, Soyez:2012hv, Krohn:2013lba, Dasgupta:2013ihk, Dasgupta:2013via, Cacciari:2014jta, Cacciari:2014gra, Bertolini:2014bba, Larkoski:2014wba}, and to measure properties of the partons initiating jets~\cite{Gallicchio:2010sw, Gallicchio:2011xc, Gallicchio:2011xq, Gallicchio:2012ez, Krohn:2012fg, Larkoski:2014pca}.  See Refs.~\cite{Abdesselam:2010pt, Altheimer:2012mn, Altheimer:2013yza} for an overview. To this toolkit the authors  recently added Qjets~\cite{Ellis:2012sn}.  In the discussion that follows we are specifically thinking in terms of ``tagging'' jets as either containing (the decay products of) a heavy boosted object
(the signal), or as being an ordinary QCD-jet (the background).

The motivation behind Qjets comes from the observation that as jets are produced through a stochastic process there is an inherent ambiguity in their reconstruction.  That is, even with a perfect algorithm one could never hope to unambiguously associate each hadron to an individual jet -- instead one typically makes a best-guess assignment using a well motivated procedure.  This is not ideal as it removes all information about the ambiguity in jet processing and tagging; any two jets that pass a set of selection cuts are assigned the same weight, even if one is unambiguously signal-like and the other is only marginally so. To address this concern the Qjets procedure processes and tags a jet using a range of plausible algorithms and grooming procedures, assigning a \textit{distribution} of possible properties to each jet.  The initial Qjets description~\cite{Ellis:2012sn} presented two central ideas: (i) a new observable volatility that characterizes the \textit{width} of the mass distribution generated by the Qjets procedure and can help distinguish jets arising from boosted heavy objects from QCD jets; and (ii) the use of the Qjets distributions to improve the statistical stability of the measurements of jet observables.   The former is an intuitively reasonable result in the sense that one expects that a jet with an underlying mass scale (\textit{i.e.}, the mass of the heavy object) will exhibit a jet mass that is more robust under changes in the details of the jet algorithm and grooming procedure compared to a background QCD jet. Volatility as a discriminating variable has recently been validated~\cite{ATLAS-CONF-2013-087, CMS-PAS-JME-13-006}  by both the ATLAS and the CMS collaborations of the LHC. The Qjets improvement in the statistical behavior of jet measurements is less intuitive and the current work has the goal of explaining the how and why of this statistical improvement.  

In order to explain why the Qjets procedure is associated with non-standard statistical analyses, let us first distinguish it from a conventional, or ``classical'' approach,
in which a jet is first groomed and then tagged to be a signal jet if its groomed mass falls within a pre-defined signal-mass window.  Such a conventional approach therefore assigns both a groomed mass $\mu_j^\text{C}$ and a tagging efficiency $\tau_j^\text{C}$ to each jet $j$.  The conventional tagging efficiency is 
a \textit{binary} tagging variable, which takes the value $1$, if the mass of the jet is within the mass-window ($\Omega$), $\mu_j^\text{C} \in \Omega$, and takes the value $0$ if the mass of the jet is not in the window, $\mu_j^\text{C} \notin \Omega$. For Qjets there is a well defined procedure (reviewed in more detail in Section~\ref{sec:review}) to groom an individual jet in a variety of ways leading a \textit{distribution} of groomed masses.  The corresponding Qjets  tagging efficiency $\tau_j^\text{Q}$ is the fraction of those masses that fall within the mass-window, while the Qjets measure of the jet mass $\mu_j^\text{Q}$ is the average of the masses that fall within the mass-window.  The fundamental difference in the statistical analysis of the Qjets case arises from the fact that  $\tau_j^\text{Q}$ exhibits a \textit{continuous} range of values in the interval $\left[0,1 \right]$, in contrast to the binary values of $\tau_j^\text{C}$.

To illustrate the unconventional features of a continuous weight $\tau^\text{Q}_j$ more specifically, consider the goal of identifying boosted $W$-jets.  A binary  $\tau_j^\text{C}$ implies a jet is \textit{either} $W$-like or  QCD-like, whereas a continuous $\tau_j^\text{Q}$ allows a jet to be treated as partially $W$-like \textit{and} partially QCD-like. Now consider an example where in an experiment the conventional approach identifies two jets with masses  $\mu_1^\text{C} = 80\gev$ and $\mu_2^\text{C} = 85\gev$ with $\tau_1^\text{C} = \tau_2^\text{C} = 1$.  One therefore reports that the experiment sees $2$ tagged $W$-jets and measures the masses of the tagged jets to be $\left( 80 + 85 \right)/2\gev = 82.5\gev$.   Contrast that result with the Qjets procedure that might assign these two jets  the same masses as the conventional approach (\textit{i.e.}, $\mu_j^\text{Q} = \mu_j^\text{C} $), but finds one jet to be more $W$-like than the other (say, $\tau_1^\text{Q} = 0.9$ and 
$\tau_2^\text{Q} = 0.2$).  So, using the Qjets procedure, the experiment instead finds $\left( 0.9 + 0.2 \right) = 1.1$ $W$-jets, and measures the $W$-mass to be $( 0.9 \times 80 +0.2 \times 85)/(0.9 + 0.2)\gev = 80.9\gev$.
Furthermore, as we explain below, both of these observables (the number of tagged jets and the measured mass from the tagged jets) are statistically more robust in the case of the Qjets procedure than in the conventional approach.  In fact, one can make a definite statement:
\begin{equation}
	\left(\frac{\delta N_T}{N_T}\right)^\text{C} = \frac{1}{\sqrt{ \epsilon N}}\, 
		\qquad \text{and} \qquad
	\frac{1}{\sqrt{  N}}\leq \left(\frac{\delta N_T}{N_T}\right)^\text{Q}  \leq \frac{1}{\sqrt{ \epsilon N}} \, ,
\label{eq:tag0}
\end{equation}
where $N_T$ represents the number of  tagged jets that arise from a physical process expected to yield  $N$ total jets and $\epsilon$ represents the efficiency of the conventional tagging procedure, $\epsilon = N_T/N$.  So, if a process is expected to yield $ N =100$ jets reconstructed at $\epsilon=50\% $ efficiency, one expects unweighted measurements of the cross section to have a statistical uncertainty of $14\%$ ($=1/\sqrt{50}$).  On the other hand, if one employs the Qjets procedure with the average tagging efficiency $\epsilon$ still 
at $50\% $,  one can achieve an uncertainty somewhere between $10\%$ and $14\%$.   {\it Thus, with Qjets  one can hope to obtain more precise results using the same data}.

More specifically, the claims in Ref.~\cite{Ellis:2012sn} regarding the uncertainties of various measurements  can be stated as 
\begin{equation}
\frac{S^\text{Q} /  \delta B^\text{Q}}{S^\text{C}/   \delta B^\text{C}} > 1 \quad \text{and} \quad 
\frac{\delta m^\text{Q} / m^\text{Q}  }{\delta m^\text{C} / m^\text{C} } < 1 \, .
\label{eq:1}
\end{equation}
These expressions use the definitions (to be explained in more detail later):
\begin{description}
\item{$S^{\text{Q}/\text{C}} =  \sum\limits_{j\in  \text{signal}} \tau_j^{\text{Q}/\text{C}} $:} Total number of signal jets correctly tagged in an experiment.
\item {$B^{\text{Q}/\text{C}} =  \sum\limits_{j\in  \text{bkg}} \tau_j^{\text{Q}/\text{C}} $:} Total number of QCD jets incorrectly tagged in an experiment.
\item {$ m^{\text{Q}/\text{C}} = \frac{ \sum\limits_{j} \mu_j^{\text{Q}/\text{C}}\tau_j^{\text{Q}/\text{C}} }
{  \sum\limits_{j} \tau_j^{\text{Q}/\text{C}} }$:} The (average) mass of the tagged jets as measured in an experiment. 
\item{$\delta B^{\text{Q}/\text{C}},\delta m^{\text{Q}/\text{C}}$:} The fluctuations in the corresponding measurements.
\end{description}
In the phenomenological studies presented below we will confirm the inequalities in Eq.~\eqref{eq:1} and attempt to provide intuitive explanations of why they hold.  Note that the explanation is not as straightforward as for Eq.~\eqref{eq:tag0}.

It is helpful to think in terms of \textit{two} types of effects contributing to the fact that the left-hand sides in Eq.~\eqref{eq:1} 
are different from $1$. As described above, an essential difference of the Qjets procedure is the shift from the binary tagging efficiency of the conventional approach, 
$\tau_j^\text{C}$ ($=0$ \textit{or} $1$),
to the continuously valued $\tau_j^\text{Q}$ ($0\leq \tau_j^\text{Q} \leq 1$). Thus jets with $\tau_j^\text{C} = 1$ can have $\tau_j^\text{Q} < 1$, while jets with $\tau_j^\text{C} = 0$, which make no contribution to the conventional analysis, can have  $\tau_j^\text{Q} > 0$ and contribute to the Qjets analysis.
These changes impact both the counting of jets and the values of weighted averages, as in the weighted average mass defined just above. One of the important results of the Qjets analysis described below is that the distribution of jet-masses assigned by the Qjets procedure ($\mu_j^\text{Q}$)  for $W$-jets is found to be more sharply peaked around $M_W$ than the $\mu_j^\text{C}$ distribution. The Qjets procedure, since it samples a variety of pruning scenarios, can include scenarios that remove unwanted radiation from a $W$-jet more 
effectively than the single conventional pruning scenario~\cite{Ellis:2009su, Ellis:2009me}.  Since it is exactly these more effective scenarios that lead to larger weights in the Qjets analysis, the resulting weighted average mass tends to be  closer to the physical $W$-mass. Thus the Qjets procedure can provide a better ``groomer'' than the classical pruning~\cite{Ellis:2009su, Ellis:2009me}.  In summary,  the improvement 
indicated in Eq.~\eqref{eq:1} stems from both the ``purely statistical'' enhancement inherent in the shift from the binomial distribution of $\tau^\text{C}_j$ to the continuous distribution of $\tau^\text{Q}_j$, which we label the ``statistical'' effect, and from the possible improvement in the measured signal mass distribution inherent in the shift of mass observable from $\mu_j^\text{C}$ to $\mu_j^\text{Q}$,  which we label the ``physics'' effect. 

Of course, the ``statistical'' and ``physics'' effects are not explicitly independent.  In an effort to a provide a quantitative separation of these two effects, we define a third, hybrid pair of variables,   ($\mu_j^\text{Q}, \tilde{\tau}_j^{{\text{Q}}}$), where the mass variable remains the same as for the usual Qjets procedure, 
but the tagging probability variable $\tilde{\tau}_j^{{Q}} $ follows a binomial distribution (similar to $\tau_j^\text{C}$) defined by
\begin{equation}
	\tilde{\tau}_j^{{\text{Q}}}  = \begin{cases} 0 & \text{ for }\tau_j^\text{Q} = 0 \\  1 & \text{ otherwise.}  \end{cases} \; 
\label{eq:tilde}
\end{equation}
With our definition of $\mu^\text{Q}_j$ in Qjets, $\tilde{\tau}_j^{{\text{Q}}}$ corresponds to tagging a jet based on whether  $\mu^\text{Q}_j$ is in the bin or not  -- \textit{i.e.} tagging efficiency is derived just like in the conventional case, but using  $\mu^\text{Q}_j$ instead of  $\mu^\text{C}_j$. Further, we define $\mu^\text{Q}_j$ such that its value is in the bin if \textit{any} of the Qjet masses for a given jet are in the bin, which is why all nonzero values for $\tau_j^\text{Q}$ yield a $\tilde{\tau}_j^{{\text{Q}}}$ value of $1$.

The left-hand sides in Eq.~\eqref{eq:1} can then be represented as \textit{products} of statistical pieces and physics pieces: 
\begin{itemize}
\item \textbf{statistical quantities:} $\frac{S^\text{Q} /  \delta B^\text{Q}}{\tilde{S}^{{\text{Q}}}  /   \delta \tilde{B}^{{\text{Q}}} } $  
and $\frac{\delta m^\text{Q} / m^\text{Q}   }{\delta \tilde{m}^{{\text{Q}}} / \tilde{m}^{{\text{Q}}}}$, 
exhibiting the impact of using a continuous versus binary variable, $\tau_j^\text{Q}$ versus $\tilde{\tau}_j^{{\text{Q}}}$;
\item \textbf{physics quantities:} $\frac{\tilde{S}^{{\text{Q}}} /  \delta \tilde{B}^{{\text{Q}}}}{S^\text{C} /   \delta B^\text{C}} $ and 
$\frac{\delta \tilde{m}^{{\text{Q}}} / \tilde{m}^{{\text{Q}}}   }{\delta m^\text{C} / m^\text{C}}$, 
primarily exhibiting the impact of the differing distributions for the mass variables $\mu_j^\text{C}$ versus $\mu_j^\text{Q}$.
\end{itemize}

The present article aims to clarify these points by presenting an explicit framework for calculating the statistics of jets obtained from the Qjets procedure, as
applied to a jet-tagging analysis. The paper is structured as follows: 
 in Section~\ref{sec:statuncer} we introduce a  statistical formalism for evaluating the uncertainties associated  with the measurement of cross-section and mass 
for a tagging efficiency described by a continuous variable,
in Section~\ref{sec:review} we review the Qjets procedure and discuss, in particular, how it leads to a mass and a tagging efficiency for a given jet,
in Section~\ref{sec:pheno} we apply the formalism derived in Section~\ref{sec:statuncer} to estimate the statistical and physics quantities outlined above, 
in Section~\ref{sec:understanding}  and Section~\ref{sec:expl} we present simple phenomenological pictures to assist in the understanding of the results for the
statistical (Section~\ref{sec:understanding}) and physics (Section~\ref{sec:expl}) effects presented in Section~\ref{sec:pheno},
and  in Section~\ref{sec:conclusion} we provide concluding remarks.  
A validation of our analytical results, derived in Section~\ref{sec:statuncer}, using Monte Carlo pseudo-experiments is provided in Appendix~\ref{sec:validation}, and more mathematical details are included in Appendix~\ref{sec:jetmass}.

%------------------------------------------------------------------------------------------------------------------
\section{\label{sec:statuncer}Statistical Uncertainties}

In this section we lay out the mathematical foundation needed to understand the statistical fluctuations of measurements when using the Qjets procedure ({\it i.e.}, non-binary tagging). This analysis applies to both signal and background measurements.

One can think of the statistical uncertainties in jet-based measurements as arising from two sources: (1) Poisson uncertainty, and (2) sampling uncertainty:

\begin{itemize}
\item \textit{Poisson uncertainty}  refers to the uncertainty in the number of events (or jets) of a certain variety produced by a process yielding discrete 
counts at some continuous rate.   For example, if a collider is expected to yield on average $N$ events (of the given variety) with a given luminosity 
($N = \mathcal{L} \sigma$, where $\sigma$ is the production cross section for this kind of event or jet) 
then the probability of it producing $n$ events is given by the Poisson distribution:
\begin{equation}
\label{eq:pois}
\text{Pois}(n | N) \equiv \frac{e^{-N} N^n }{ n!}\,,  \quad \langle n \rangle_\text{Pois} = N\,, \quad  
\sigma^2_\text{Pois} 
= N \,.
\end{equation}
Thus the variance ($\sigma^2_\text{Pois}$) of this distribution is $N$ as indicated, which tells us that the characteristic size of the variation in the number of events (of the given variety) produced with a given luminosity  from one experimental run to the next  is $\sqrt{N}$.  

\item \textit{Sampling uncertainty} refers to the uncertainty in the way the events will be reconstructed by the analysis procedure, leading to fluctuations in the tagging 
rate sample-to-sample. Let us illustrate this point with an explicit example. Consider that we are trying to identify jets containing $W$ decays with an algorithm characterized by a 
given tagging efficiency (say $70\%$).
 By sampling uncertainty we refer to the fact that for one sample of $100$ signal jets the procedure might tag $75$ jets as $W$-like, while for another 
sample of $100$ signal jets it might only tag $65$.

\end{itemize}

\begin{figure}[!ht]
	\centering 
		{\includegraphics[width=0.80\textwidth]{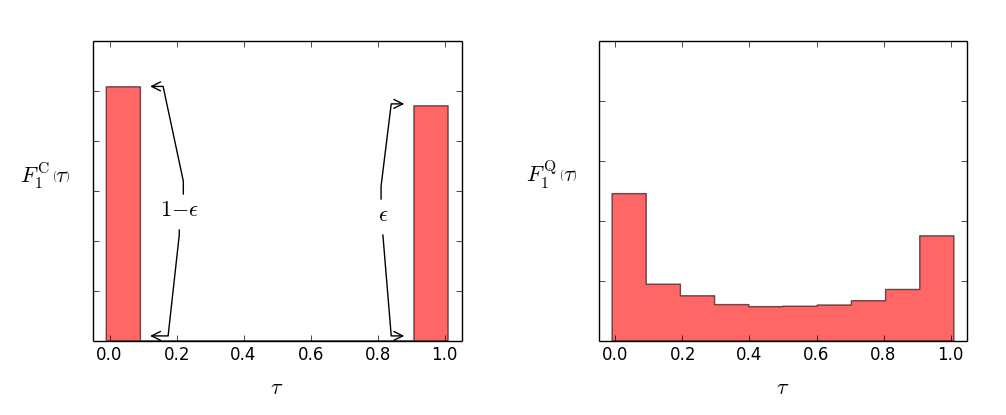} \label{fig:WF1Q} }
\caption{Illustration of how  $F_1^{\text{C}}(\tau)$ (left) and $F_1^{\text{Q}}(\tau)$ (right) may look for a sample of $W$-jets. Note the binomial nature of $F_1^{\text{C}}(\tau)$  as opposed to the continuous distribution of $\tau$ in  $F_1^{\text{Q}}(\tau)$.}
\label{fig:1}
\end{figure}

The next step is to explain why the probability distribution describing the tagging of jets in the Qjets procedure is fundamentally different from the conventional procedure, resulting in qualitatively (and quantitatively) different expressions for the sampling uncertainty, as well as the total statistical uncertainty. 
Recall that a conventional tagging procedure assigns a binary valued weight $\tau$ of either $1$ or $0$ (\textit{i.e.}, tagged or not-tagged) to a jet.  Such a procedure is usually characterized by a tagging efficiency $\epsilon$, which means that, on average, a fraction $\epsilon$ of jets selected at 
random from a sample of $W$-jets
will be tagged.  Thus the explicit probability distribution function (or pdf) for tagging $1$-jet, picked at random from a set of $W$-jets, by a conventional ($\text{C}$) 
procedure can be simply represented as: 
\begin{equation}
	\label{eq:convF1}
	F_1^{\text{C}}(\tau) = (1-\epsilon)\delta(\tau)+\epsilon\delta(\tau-1) \,, 
\end{equation}
(where $\delta(\tau)$ is the usual delta function that vanishes everywhere except at $\tau=0$, but is sufficiently singular at $\tau = 0$ to satisfy
$\int d\tau f(\tau) \delta(\tau) = f(0)$ for any range of integration that includes the origin). 
This form is illustrated in the left-hand plot in  Figure~\ref{fig:1}.  In contrast,
the weight  $\tau$ assigned by a Qjets procedure can have any value in the interval $\left[0,1\right]$.  We label the pdf for tagging $1$-jet 
(picked at random from a set of $W$-jets)  with probability $\tau$ by the Qjets procedure as $F_1^{\text{Q}}(\tau)$.  Note that, unlike Eq.~\eqref{eq:convF1}, $F_1^{\text{Q}}(\tau)$ is a continuous function of $\tau$, as illustrated in the right-hand plot in Figure~\ref{fig:1}.  

These $1$-jet tagging probability distribution functions (the $F_1(\tau)$'s illustrated in Figure~\ref{fig:1}) are central to our analysis. As we will show later, the statistical uncertainties associated with various tagged $1$-jet measurements are given entirely in terms of the first few moments of the $F_1(\tau)$.  In particular, we define the average and variance (and the normalization) of $F_1(\tau)$ to be:
\begin{equation}
\label{eq:moments}
\langle \tau \rangle = \int_0^1 \tau F_1(\tau) d\tau\,, \quad \text{ and } \qquad
\sigma_\tau^2 = \int_0^1 (\tau -\langle \tau \rangle)^2 F_1(\tau) d\tau \, , \qquad \left( \int_0^1  F_1(\tau) d\tau=1\right) \,.
\end{equation}
Note that, in the special case of the conventional procedure as in Eq.~\eqref{eq:convF1},
\begin{equation}
\label{eq:momentsconv}
\langle \tau^\text{C} \rangle = \int_0^1 \tau F_1^\text{C}(\tau) d\tau \ = \epsilon \, , \quad \text{ and } \qquad
(\sigma_\tau^\text{C})^2 = \int_0^1 (\tau -\langle \tau \rangle)^2 F_1^\text{C}(\tau) d\tau \ =   \epsilon (1-  \epsilon)\,,
\end{equation}
where, as above, $\epsilon$ is the average tagging efficiency in the conventional procedure.  The results of Eq.~\eqref{eq:momentsconv} 
are just what we expect from the binomial distribution corresponding to a \textit{binary} valued weight.  Note that the differences between the two
distributions in Figure~\ref{fig:1} (binary versus continuous, where the latter has more support near the average value) 
already suggest that $\sigma_\tau^\text{Q} < \sigma_\tau^\text{C}$,
even for cases where $\langle \tau^\text{Q} \rangle \approx \epsilon$.

For the corresponding hybrid analysis of Eq.~\eqref{eq:tilde} we have a distribution similar to Eq.~\eqref{eq:convF1},
\begin{equation}
	\label{eq:tildeF1}
	\widetilde{F}_1^{\text{Q}}(\tilde\tau) = (1-\tilde{\epsilon})\delta(\tilde\tau)+\tilde{\epsilon}\delta(\tilde\tau-1) \,, 
\end{equation}
with moments
\begin{equation}
\label{eq:momentstilde}
\langle {\tilde\tau}^\text{Q} \rangle =  \tilde\epsilon , \quad \text{ and } \qquad
(\sigma_{\tilde\tau}^\text{Q})^2  =   \tilde\epsilon (1-  \tilde\epsilon)\, .
\end{equation}
Note that, since the jets that were tagged in the conventional analysis are typically still tagged, and the Qjets procedure allows more jets to be tagged, with $\tilde\tau_j^\text{Q} = 1$
in the hybrid analysis, we expect that $\tilde\epsilon > \epsilon$.  
%---------------------------------------------------
\subsection{\label{subsec:csuncer}Cross-section measurement}

As a first detailed example consider the statistical uncertainties inherent in the measurement of the production cross-section of jets containing the desired heavy particle.
First, imagine that $N_S$ jets are selected at random from a set of $W$-jets. The total number of (correctly) tagged $W$-jets (or $N_T$) is then given as 
\begin{equation}
\label{eq:nr}
N_{T} = \sum_{ j=1}^{N_S} \tau_j \, .
\end{equation}
Since $N_T$ is a sum of weights, it can exhibit non-integral values for the Qjets procedure. 
The probability distribution describing $N_T$, for a given sample size $N_S$, can be constructed in terms of $F_1$,
\begin{equation}
\label{eq:F1NS}
F_{N_S}(N_T)  =\left[  \prod_{k=1}^{N_S}\int F_1(\tau_k) d\tau_k \right] \delta \left(N_T - \sum_{k=1}^{N_S} \tau_k\right)\,.
\end{equation}
For future reference the first two moments of this general distribution are
\begin{equation}
\label{eq:F1NSNT}
\langle N_T \rangle_{N_S} = \int N_T dN_T F_{N_S}(N_T)  = 
 \left[\prod_{k=1}^{N_S}\int F_1(\tau_k) d\tau_k \right] \sum_{k=1}^{N_S} \tau_k = N_S \langle \tau \rangle \,,
\end{equation} 
and
\begin{equation}
\label{eq:F1NSNT2}
\begin{split}
\langle N^2_T \rangle_{N_S} & =  \int N_T^2 dN_T F_{N_S}(N_T)  
=  \left[\prod_{k=1}^{N_S}\int F_1(\tau_k) d\tau_k \right] \left(\sum_{k=1}^{N_S} \tau_k\right)^2  \\
& =  \left[\prod_{k=1}^{N_S}\int F_1(\tau_k) d\tau_k \right] \left(\sum_{k=1}^{N_S} \tau_k^2 + \sum_{k\neq l}^{N_S} \tau_k \tau_l\right) \\
& = N_S \langle \tau^2 \rangle +N_S(N_S-1) \langle \tau \rangle^2 =  N_S^2 \langle \tau \rangle^2+ N_S \left( \langle \tau^2 \rangle - \langle \tau \rangle^2 \right)  
\equiv N_S^2 \langle \tau \rangle^2 +N_S \sigma_\tau^2 \,.
\end{split}
\end{equation} 

For the conventional procedure (with a binary valued $\tau$) 
$F_{N_S}^\text{C}(N_T)$ is given by 
the probability of selecting $N_T$ objects from a set of $N_S$ objects and the pdf is given by a \textit{Binomial} distribution of mean $\epsilon$:
\begin{equation}
\label{eq:convF1NS}
\begin{split}
F_{N_S}^\text{C}(N_T) & \ = \ \left[  \prod_{k=1}^{N_S}\int F_1^\text{C}(\tau_k) d\tau_k \right] \delta \left(N_T - \sum_{k=1}^{N_S} \tau_k\right) \\
& \ = \ \frac{N_S!}{N_T!(N_S-N_T)!}\epsilon^{N_T}(1-\epsilon)^{N_S-N_T} 
\  \equiv \  \text{B}(N_T | N_S , \epsilon) \,,
\end{split}
\end{equation}
with moments
\begin{equation}
\begin{split}
\langle N^\text{C}_T \rangle_{N_S} & =  N_S \ \epsilon  \\
\langle \left( N^{\text{C}}_T \right)^2 \rangle_{N_S} & =  N_S^2 \ \epsilon^2 + N_S\ \epsilon (1-\epsilon) \,.
\end{split}
\end{equation}

Next we consider measuring the production cross section for the tagged jets. As noted above, the total statistical uncertainty depends on both the 
Poisson uncertainty and the sampling uncertainty.  If the expected number of jets (for a given luminosity $\cal L$) is 
$N$, on average the probability $\text{P}$ of tagging $N_T$ jets is given by:
\begin{equation}
\label{eq:eventprob}
	\text{P}(N_T | N) = \sum_{N_S=N_T}^\infty \text{Pois}(N_S | N)\times F_{N_S}(N_T)\,.
\end{equation}
Evaluating Eq.~\eqref{eq:eventprob} in the conventional case is easier than one might expect, since the combination of a Poisson process 
and a Binomial process is still a Poisson process.
We have
\begin{equation}
\label{eq:csconv}
\begin{split}
\text{P}^\text{C}(N_T | N) &\ = \ \sum_{N_S=N_T}^\infty \text{Pois}(N_S|N)\times F_{N_S}^\text{C} (N_T) \\
 & \ = \  \sum_{N_S=N_T}^\infty \text{Pois}(N_S|N)\times \text{B}(N_T | N_S,\epsilon)
\ = \  \text{Pois}(N_T | \epsilon N),
\end{split}
\end{equation}
{\it i.e.}, it is a Poisson distribution with mean $\epsilon N$.  Thus we can still apply our ``$\sqrt{N}$'' intuition.  Using Eq.~\eqref{eq:csconv} 
 (and Eq.~\eqref{eq:pois}) we find that the fractional uncertainty in the number of conventionally tagged jets is
\begin{equation}
\label{eq:dnnrconv}
\frac{\delta N_T^\text{C}}{N_T^\text{C}} = \frac{\sqrt{\sigma^2_\text{Pois}(N_T)}}{\langle N_T \rangle_\text{Pois}}=\frac{\sqrt{ \epsilon N}}{\epsilon N}  
= \frac{1}{\sqrt{ \epsilon N}}\,,
\end{equation}
as already noted in Eq.~\eqref{eq:tag0}.

Thus, if we observe $100$ events with tagged signal jets in ${\cal L}=1~{\rm fb}^{-1}$ with $\epsilon=50\%$, we would report a cross section for signal jets of $200 \pm 20~{\rm fb}$ (\textit{i.e.}, $\sigma = N_T/\epsilon/{\cal L}=100/0.5$ fb,  and $\delta \sigma/\sigma = \delta N_T/N_T=1/\sqrt{100} = 1/10$).

Evaluating statistical uncertainties for a general $F_1(\tau)$, \textit{e.g.}, a Qjets $F_1^\text{Q}(\tau)$, is slightly more complicated.  
In particular, for the Qjets case $N_T$ is a sum of 
non-integer weights and so can exhibit non-integer values.  For example, consider a sample of 5 jets/events.  If, at the non-integer value $4.5$, 
$F^\text{Q}_5(4.5) = 0.1$, then we interpret this
to mean that the probability of measuring one jet/event in the bin $4.5\pm \rho/2$ is  $0.1\times\rho$, for infinitesimal $\rho$.  In the following
manipulations we treat $N_T$ as a continuous variable. The mean of the distribution $\text{P}(N_T | N)$  is obtained from (recall Eq.~\eqref{eq:F1NSNT})
\begin{equation}
\begin{split}
\label{eq:eventave}
\langle N_T \rangle & = \int N_T\, dN_T\,\text{P}(N_T | N) =  \sum_{N_S=0}^\infty \text{Pois}(N_S|N) \int_0^{N_S} N_T\, dN_T\, F_{N_S}(N_T)  \\
                                     & =  \sum_{N_S=0}^\infty \text{Pois}(N_S|N) N_S \langle \tau \rangle = \langle \tau \rangle N \, .
\end{split}                
\end{equation}
The second moment of $\text{P}(N_T | N)$  is found from (recall Eq.~\eqref{eq:F1NSNT2})
\begin{equation}
\begin{split}
\label{eq:eventave2}
\langle N_T^2 \rangle  & = \int N_T^2\, dN_T\,\text{P}(N_T | N) 
=  \sum_{N_S=0}^\infty \text{Pois}(N_S|N) \int_0^{N_S} N_T^2\, dN_T\, F_{N_S}(N_T) \notag \\
                                     & =  \sum_{N_S=0}^\infty \text{Pois}(N_S|N) \left( N_S \sigma_\tau^2 +N_S^2  \langle \tau \rangle^2\right)  \\
                                     & =   N \sigma_\tau^2 + N(N+1) \langle \tau \rangle^2 \,.
\end{split}
\end{equation}
So the desired variance is
\begin{equation}
\begin{split}
\label{eq:eventvar}
(\delta N_T)^2 \ \equiv \ \langle N_T^2 \rangle - \langle N_T \rangle^2 & =  N \sigma_\tau^2 + N(N+1) \langle \tau \rangle^2 - N^2  \langle \tau \rangle^2    \\
                                     & =   N \left( \sigma_\tau^2 +  \langle \tau \rangle^2 \right) \,.
\end{split}
\end{equation}

This is the general result including the analysis above for the conventional case in Eq.~\eqref{eq:dnnrconv}, when we recall that in the conventional scenario
(as in Eq.~\eqref{eq:momentsconv})
$\langle \tau^\text{C} \rangle = \epsilon$, $(\sigma_\tau^\text{C})^2 = \epsilon (1-\epsilon)$ so that  
$(\sigma_\tau^\text{C})^2  + \langle \tau^\text{C} \rangle^2  = \epsilon$.   In the Qjets analysis the distribution $F_1(\tau)$ 
becomes non-zero at intermediate $\tau$ values ($\tau \neq 0,1$), which, as already suggested, serves to \textit{reduce} $\sigma_\tau$ 
from its ``conventional'' value, as we will see explicitly shortly.

So it follows that for a general probability distribution $F_1(\tau)$ we have
\begin{equation}
\label{eq:dnnr}
\frac{\delta N_T}{N_T}=\frac{1}{\sqrt{N}}\times\sqrt{1+\frac{\sigma_\tau^2}{\langle \tau \rangle^2}}\,.
\end{equation}
Since in the general case, $\tau_k \leq 1.0$ and thus $\tau^2_k \leq \tau_k$, the averages must obey
\begin{equation}
\label{eq:fluc}
\langle \tau^2 \rangle \leq \langle \tau \rangle 
	\quad \Rightarrow \quad 
		\sigma_\tau^2 \equiv \langle \tau^2 \rangle - \langle \tau \rangle^2 \leq \langle \tau \rangle(1 - \langle \tau \rangle)\,.
\end{equation}
Thus we obtain 
(essentially as claimed in the Introduction) that
\begin{equation}
\label{eq:dcsbound}
\frac{1}{\sqrt{  N}}\ \leq \ \frac{\delta N_T}{N_T} \ \leq \ \frac{1}{\sqrt{ \langle \tau \rangle N}}.
\end{equation}
Comparing this to Eq.~\eqref{eq:dnnrconv} we see that the \textit{upper} limit is saturated for the conventional procedure with binary valued tagging.  
This allows for the the fractional uncertainty in the cross-section measurement to be reduced by up to a factor of 
$\sqrt{\langle \tau^\text{C} \rangle}$ ($=\sqrt{\epsilon}$) if weighted jets are used in the measurement.  
{\it This is the advantage of using weighted jets -- while we are still subject to the Poisson uncertainties in Eq.~\eqref{eq:csconv}, 
the sampling uncertainties, encoded in $\text{B}(N_T | N_S ,\epsilon)$ for a conventional tagging procedure, are reduced.}

%---------------------------------------------------
\subsection{\label{subsec:massuncer}Mass measurement}

The statistical uncertainty of a cross section measurement is straightforward to compute with Qjets because the probability distribution for the 
number of tagged events factorizes nicely into one factor capturing the effects of Poisson uncertainties and one capturing the effects of sampling uncertainties 
(see  Eq.~\eqref{eq:eventprob}).  This is not generally true for other quantities that involve a weighted average rather than a simple sum,\textit{e.g.}, the average tagged jet mass is defined by
\begin{equation}	
\label{eq:mr}	 
m _T= \frac{ \sum_{ j=1}^{N_S} \mu_j \tau_j}{\sum_{ j=1}^{N_S} \tau_j }= \frac{1}{N_{T} } \sum_{ j=1}^{N_S} \mu_j \tau_j \,.
\end{equation}
The corresponding expression relevant to the hybrid analysis of Eq.~\eqref{eq:tilde} is
\begin{equation}	
\label{eq:mrtilde}	 
\widetilde{m}_T= \frac{ \sum_{ j=1}^{N_S} \mu_j \tilde{\tau}_j}{\sum_{ j=1}^{N_S} \tilde{\tau}_j }= \frac{1}{\tilde{N}_{T} } \sum_{ j=1}^{N_S} \mu_j \tilde{\tau}_j \,.
\end{equation}

One can still relate the relevant uncertainties to the underlying probability distribution functions; however, the resulting 
expressions are more complicated.  In particular, $F_1(\tau)$ is no longer enough. We now need to to
know the probability distribution as a function of \textit{both} $\tau$ and $\mu$.  We label this
distribution $F_1(\mu, \tau)$, which denotes the probability distribution in the $\left(\mu, \tau \right)$ plane. 
Note that $F_1(\tau)$ is simply related to $F_1(\mu, \tau)$ by
\begin{equation}
\label{eq:tauonly}
F_1(\tau)  =  \int d\mu F_1 (\mu,\tau) \; .
\end{equation}

\begin{figure}[!ht]
	\centering 
	{\includegraphics[width=0.80\textwidth]{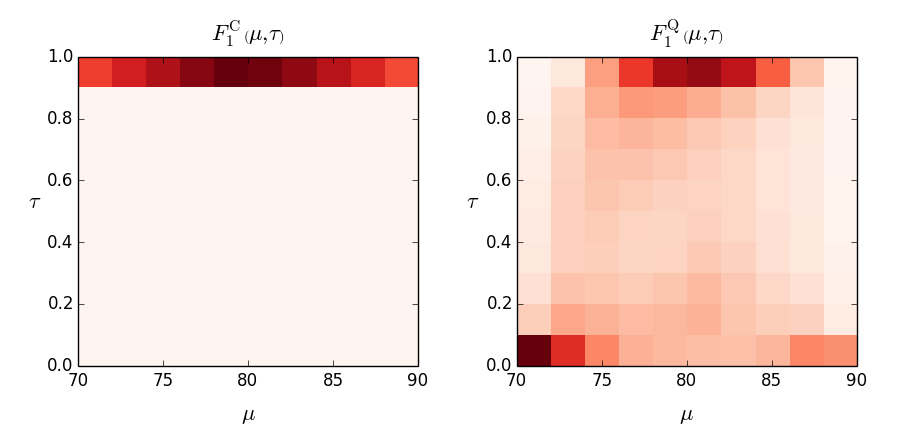} \label{fig:WF1MTQ} }
\caption{Illustration of how  $F_1^{\text{C}}(\mu,\tau)$ and $F_1^{\text{Q}}(\mu,\tau)$ may behave for a sample of $W$-jets. Since these plots are for illustration purposes only, we do not provide the numerical values associated with different shades of red. Qualitatively, the lightest shade in these plots represents a vanishing $F_1(\mu,\tau)$ value, and a darker shade represents a larger value of  $F_1$.
Note that all jets in the jet mass window $\left(70-90\right)\gev$ are tagged  (with $\tau = 1$) in a conventional procedure and so only the $\tau = 1$ boxes will be non-zero for $F_1^{\text{C}}(\mu,\tau)$. 
On the other hand, $F_1^{\text{Q}}(\mu, \tau)$ can be non-zero in the entire $\left(\mu,\tau \right)$ plane. 
}
\label{fig:2}
\end{figure} 
For illustration we show the $F_1^\text{C} (\mu,\tau)$ and $F_1^\text{Q} (\mu,\tau)$ distributions in Figure~\ref{fig:2} derived from a sample of $W$-jets.  In the conventional procedure, a jet with jet mass inside a pre-defined mass window,
for example, $\Omega = \left(70-90\right)\gev$ for $W$-tagging, is tagged (with $\tau = 1$). This fact is demonstrated by $F_1^\text{C} (\mu,\tau)$, where all non-zero entries are in the bin at $\tau = 1$ and the jet mass distribution peaks 
around the $W$-mass. 
On the other hand, $F_1^\text{Q} (\mu,\tau)$ shows that there are non-zero probabilities for tagging jets with efficiency $\tau^\text{Q}$ in the 
full range $[0,1]$ for jet masses in the tagging window $\Omega$.  Note that the contributions that lead to the
strictly $\tau = 0$ part of the distribution (see, for example, Eq.~\eqref{eq:convF1}) all arise from $\mu$ values \textit{outside} of $\Omega$. 

In this section, we simply define moments of the two-dimensional distribution, and leave all technical details to Appendix~\ref{sec:jetmass}.  
The moments of interest, the single averages, the two-dimensional mean,  variance and covariance are given by
\begin{equation}
\label{eq:moments2}
\begin{split}
\langle \tau \rangle & \equiv \int d\mu \int_0^1 d\tau\, \tau  F_1(\mu,\tau) 
=  \int_0^1 d\tau\, \tau  F_1(\tau)  \,, \\
\langle \mu\tau \rangle & \equiv \int d\mu \int_0^1 d\tau\, \mu \tau  F_1(\mu,\tau) 
\,, \\
\sigma^2_{\tau} &\equiv \int d\mu \int_0^1 d\tau\, (\tau - \langle \tau\rangle)^2  F_1(\mu,\tau)  \,, \\
\sigma^2_{\mu\tau} &\equiv \int d\mu \int_0^1 d\tau\, (\mu \tau - \langle \mu\tau\rangle)^2  F_1(\mu,\tau)  \,, \\
\sigma(\tau,\mu\tau) & \equiv  \int d\mu \int_0^1 d\tau\, (\mu\tau  - \langle \mu \tau \rangle)(\tau - \langle \tau \rangle)  F_1(\mu,\tau)  \, .
\end{split}
\end{equation} 
Note especially that, since $\tau$ and $\mu$ are correlated by $F_1(\mu,\tau)$, we are now interested in both the variance of the parameters $\tau$ and $\mu\tau$ \textit{and} in the covariance $\sigma(\tau, \mu \tau)$.
 
So we are now ready to consider the measurement of the average (weighted)  jet mass as defined in Eq.~\eqref{eq:mr}, where we want to understand the expected improvement in precision from using the Qjets technique. Proceeding essentially as we did in the cross section case, the expected average value of $m_T$ in a sample of $N_S$ jets, is given by 
(recall Eq.~\eqref{eq:mr})
\begin{equation}
\label{eq:mraveNS}
\langle m_T \rangle_{N_S} \simeq \frac{ \langle \mu \tau\rangle}{\langle \tau \rangle}\left[1+ \frac{\sigma_\tau^2}{N_S \langle \tau \rangle^2} 
- \frac{\sigma(\tau,\mu\tau)}{N_S \langle \mu\tau\rangle \langle \tau\rangle}  \right]\,.
\end{equation}
As explained in the appendix, we are expanding in the fluctuations around the average values and assuming that the 
higher order fluctuations are negligible.  The corresponding variance in this quantity is given by
\begin{equation}
\label{eq:mrvarNS}
\left(  \delta m_T \right)^2_{N_S} = 
\langle (m_T - \langle m_T \rangle_{N_S})^2 \rangle_{N_S} \simeq \frac{ \langle \mu \tau\rangle^2}{N_S \langle \tau \rangle^2}\left[\frac{\sigma_{\mu\tau}^2}{\langle \mu\tau \rangle^2} 
+\frac{\sigma_{\tau}^2}{\langle \tau \rangle^2}-2 \frac{\sigma(\tau,\mu\tau)}{\langle \mu\tau\rangle \langle \tau\rangle}  \right]\,.
\end{equation}
If we average over samples (to take into account the Poisson uncertainties) within an experiment with a given luminosity, 
then we have $N_S \to N = \sigma {\cal L}$ in the denominator of both Eq.~\eqref{eq:mraveNS}) and Eq.~\eqref{eq:mrvarNS}, 
plus corrections of order $1/N^2$.  Combining the above results, the ratio of the fluctuations to the 
average value can be written as: 
\begin{equation}
\label{eq:mrvarN}
\left(  \frac{\delta m_T }{\langle m_T \rangle} \right)^2 =  \frac{ 1}{N}
\left( \frac{\sigma_{\mu\tau}^2}{\langle \mu\tau \rangle^2} 
+ \frac{\sigma_{\tau}^2}{\langle \tau \rangle^2}   - 
2  \frac{\sigma(\tau,\mu\tau)}{\langle \mu\tau\rangle \langle \tau\rangle}
\right) + \ord{\frac{ 1}{N^2}} \,.
\end{equation}

We can easily evaluate this quantity for the conventional binary tagging procedure. By definition, and as illustrated in  Figure~\ref{fig:2}, 
$\tau = 1$ for $\mu \in \Omega$ when we consider the pdf $F_1^\text{C}(\mu, \tau)$. The fractional mass uncertainty of Eq.~\eqref{eq:mrvarN} for the
conventional tagging procedure with average tagging efficiency $\epsilon$ is then
\begin{equation}
\label{eq:mrvarNconv}
\left(  \frac{\delta m_T^\text{C} }{\langle m_T^\text{C} \rangle} \right)^2 =  \frac{ 1}{N} \times 
   \frac{ \left( \sigma_{\mu}^{\text{C}} \right)^2}{\epsilon \langle \mu^\text{C} \rangle^2}   
\ + \ \ord{\frac{ 1}{N^2}} \, ,
\end{equation}
where we define the properly normalized mass distribution moments \textit{in} the mass window $\Omega$ as
\begin{equation}
\begin{split}
	\langle \mu \rangle \ &  \equiv \ \frac{1}{N_\Omega}  \  
		\int_\Omega d\mu \int_0^1 d\tau\, \mu   F_1(\mu,\tau)  \, , \\
	\sigma^2_{\mu} \  & \equiv \  \frac{1}{N_\Omega}  \  
		\int_\Omega d\mu \int_0^1 d\tau\, \left( \mu - \langle \mu \rangle \right)^2 \  
			 F_1(\mu,\tau)   \, ,  \\
	& \quad  \text{with}  \qquad 
	N_\Omega \ = \  \int_\Omega d\mu \int_0^1 d\tau\,  F_1(\mu,\tau) \, .
\end{split}	
\label{eq:omegamoments}
\end{equation}
Here $N_\Omega$ fixes the normalization of the pdf $F_1$ in the mass window  $\Omega$.
In Eq.~\eqref{eq:mrvarNconv} we follow the convention in Eq.~\eqref{eq:momentsconv} to denote that the moments $ \langle \mu^\text{C} \rangle $ and $ \sigma_{\mu}^{\text{C}} $ are calculated from  Eq.~\eqref{eq:omegamoments} using  the conventional pdf $F_1^\text{C} \left( \mu ,\tau\right)$.  
Note that in the conventional case the normalization is
\begin{equation}
	\left( N_\Omega \right)^\text{C} \ = \   \int_\Omega d\mu \int_0^1 d\tau\,  F_1^\text{C}(\mu,\tau) \ = \ 
		\int_0^1 d\tau \left(  \int_\Omega d\mu  \,  F_1^\text{C}(\mu,\tau) \right)   \ = \ 
		\int_0^1 d\tau \, \epsilon \, \delta(1-\tau)  \ = \  \epsilon \; ,
\label{eq:normclassical}
\end{equation}
where we have used the fact that in the conventional or classical analysis all jets in the tagging window have $\tau=1$. Once again,  the reader is directed to Appendix~\ref{sec:jetmass} for details.

%------------------------------------------------------------------------------------------------------------------
\section{\label{sec:review}Review of Qjets}

The purpose of this section is to demonstrate how the Qjets procedure assigns a jet mass ($\mu_j^\text{Q}$) and tagging efficiency 
($\tau_j^\text{Q}$) to a given jet $j$. Before describing the details, let us first review the general idea of the procedure. 
As suggested in Ref.~\cite{Ellis:2012sn}, we start with jets identified using a standard algorithm like Anti-$k_T$~\cite{Cacciari:2008gp}.  We recluster  
the  constituents of the given jet using a sequential and probabilistic recombination algorithm, such as $k_T$~\cite{Catani:1993hr, Ellis:1993tq} or
Cambridge/Aachen (C/A)~\cite{Dokshitzer:1997in, Wobisch:1998wt, Wobisch:2000dk}. During clustering, pruning~\cite{Ellis:2009su, Ellis:2009me} 
is performed  in order to remove unwanted elements in the jet, \textit{i.e.}, those elements not arising from the decay of the
desired heavy object.  Through pruning we map a jet to its pruned version.  
If the above set of steps is repeated on the same jet using a slightly different recombination metric as explained below (the Qjets procedure), 
we obtain a different 
four-vector after pruning due to the probabilistic nature of the Qjets clustering 
algorithm. We iterate the procedure a number of times (say $N_\text{iter}$) to map a jet to a set of pruned four-vectors. The quantities $\mu_j^\text{Q}$ and $\tau_j^\text{Q}$ are then calculated from the invariant masses of these pruned four-vectors.

In more detail, sequential recombination algorithms build up jets by merging four-momenta in pairs over many steps.  The behavior of the algorithms is 
determined by the metric for measuring the ``distance'' between  four-momenta.  At each stage in the jet clustering, one identifies the pair of four-momenta 
with the smallest distance and merges them together (\text{i.e.}, adds the corresponding 4-momenta and replaces the merged pair
with this sum 
in the updated list of 4-momenta).  This merging step is repeated on the list of 4-momenta until all remaining 4-momenta are
separated by more than a predefined cutoff.  
See Ref.~\cite{Salam:2009jx} for a more comprehensive discussion.  For instance, the $\kt$~\cite{Catani:1993hr, Ellis:1993tq} and 
$\CA$~\cite{Dokshitzer:1997in, Wobisch:1998wt, Wobisch:2000dk} algorithms correspond to the following metrics: 
\begin{equation}
		d^{\kt}_{ij}  \equiv \ \text{min}\{ p_{T_i}^2, p_{T_j}^2 \} \Delta R_{ij}^2  \; {\rm\  and\ \ } d^{\CA}_{ij}  \equiv \ \Delta R_{ij}^2 \; ,
	\label{eq:dij_defns}	
\end{equation}
where $\Delta R_{ij}^2=\Delta y_{ij}^2  + \Delta \phi_{ij}^2$ is the squared angular distance between a pair of four-momenta $i$ and $j$
(with $y$  the usual rapidity and $\phi$ the azimuthal angle).  Thus the C/A 
algorithm merges the 4-momenta in strict order of their angular separation with closest merged first.  The $k_T$ algorithm, in contrast, gives some emphasis to merging the
smallest $p_T$ elements first and so the two algorithms will tend to identify jets with slightly different constituents. 

As implemented in  Ref.~\cite{Ellis:2012sn}, the Qjets procedure also processes jets via pairwise mergings with pruning applied at each merging step.  
However, unlike traditional clustering which works deterministically,  
Qjets uses a probabilistic clustering procedure:
\begin{enumerate}
\item At every stage of clustering, for each pair of four-vectors (say, $i$ and $j$), the conventional distance metric $d_{ij}$ from Eq.~\eqref{eq:dij_defns} 
(for $k_T$ or C/A) is evaluated for all such pairs.  This is translated into a weight $\omega_{ij}^{(\alpha)}$ via
\begin{equation}
	\label{eq:weight1}
	\omega_{ij}^{(\alpha)} \equiv \exp \left\{ - \alpha  \frac{(d_{ij} - \dmin)}{\dmin}  \right\} \;,
\end{equation}
where $\dmin$ is the smallest $d_{ij}$ at this stage in the clustering process and  $\alpha$ (termed {\it rigidity}) is a continuous real parameter.  This weight
is then used to assign a probability $\Omega_{ij}$ to each pair via
\begin{equation}
\Omega_{ij} = \omega_{ij}/N, \, \text{where}\, N = \sum_{\pair{i}{j}}\omega_{ij}\, .
\label{eq:prob}
\end{equation}

\item A random number is generated and used to select a pair $\pair{i}{j}$ with probability  $\Omega_{ij}$.  Note that the conventional clustering process will
always choose the pair with the minimum $d_{ij}$ at this point and corresponds to the limit $\alpha \to +\infty$.

\item Having chosen the pair $\pair{i}{j}$, the standard pruning procedure is applied. The softer of the two selected four-momentum  pair $\pair{i}{j}$ is discarded, 
if \textit{both} of the following criteria are satisfied for a given set of parameters $\left( z_\text{cut}, D_\text{cut} \right)$.  
\begin{equation}
	z \ \equiv \ \frac{\text{min} \left( p_{T_i}, p_{T_j} \right)}{p_{T_p}} \ <  \ z_\text{cut}
		\quad \text{and} \quad  
 	\Delta R_{ij}   \ > \	 D_\text{cut}\,.	
				\label{eq:prune}
\end{equation}
Otherwise, the pair is merged. 

\item Steps ($1$-$3$) are repeated until all constituents are clustered.  The invariant mass of the resultant pruned four-vector is stored for further analysis. 

\item Steps ($1$-$4$) are repeated $N_\text{iter}$ times. This procedure yields a set of  $N_\text{iter}$ masses for every jet it operates on. 
Due to the random numbers in step 2 these masses are generally not the same, but instead define a distribution of masses.     
\end{enumerate}

\begin{figure}
	\centering
	\includegraphics[width=0.4\textwidth]{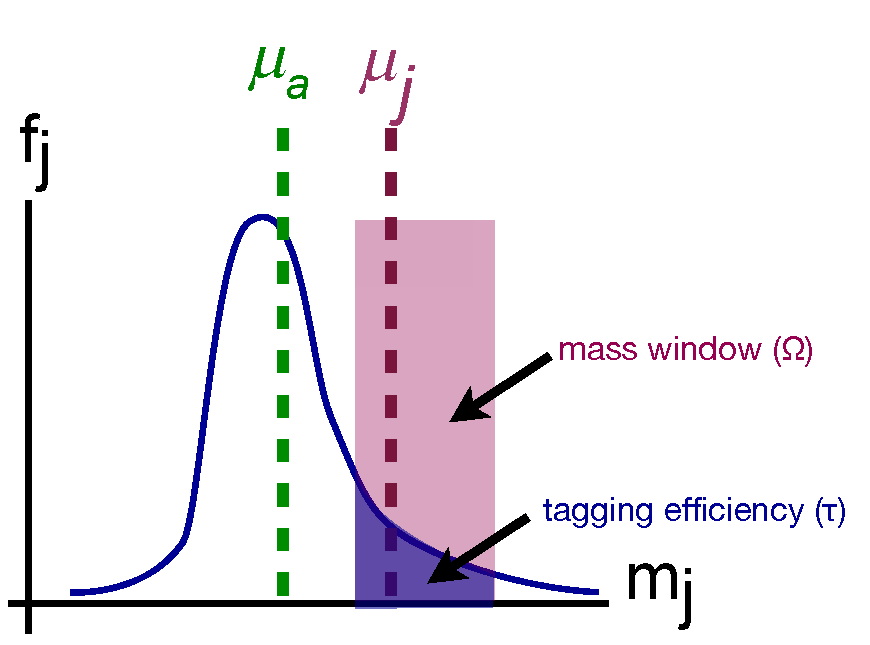}
	\caption{\label{fig:window} Sketch of the pruned jet mass distribution for a jet processed many times with Qjets. The red area represents the mass window
($\Omega$), the fraction of the jetmass distribution within the mass window (blue) is the tagging efficiency of the jet $\tau_j^\text{Q}$, and $\mu_j^\text{Q}$ is the mean jetmass-in-the-window.  $\mu_a$ is the average jet mass for the entire distribution.}
\end{figure}

In summary, the Qjets procedure maps the initial jet $j$ to a set of  masses, $\left\{m_{j,k} \right\}$, where $k$ takes integer values in $\left[1, N_\text{iter}\right]$.  
For each jet $j$ we can construct a probability distribution  $f_j(m_j)$ as suggested in  Figure~\ref{fig:window}, 
with normalization $\int f_j (m_j) dm_j= 1$.  For $N_\text{iter}\gg 1$ this distribution will be relatively smooth and we will treat it as a continuous function, 
\begin{equation}
\label{eq:jetfct}
f_j(m_j) \equiv \lim_{N_\text{iter} \gg 1} \frac{1}{N_\text{iter}}\sum_{k=1}^{N_\text{iter}} \delta (m_j-m_{k,j})\,.
\end{equation}

As described above, to define a tagging process, for example for W-jets, we focus on the W-like mass window $\Omega$ illustrated in Figure~\ref{fig:window}.  
For a given jet $j$, the tagging probability $\tau_j^\text{Q}$ is the fraction of 
the $N_\text{iter}$ clustering sequences yielding a pruned mass within the $W$-window, 
\begin{equation}
\label{eq:tau}
\tau_j^\text{Q} = \frac{1}{N_\text{iter}} \sum_{k \backepsilon\, m_{j,k} \in  \: \Omega} 1  = \int_\Omega f_j (m_j) dm_j \,.
\end{equation}
Similarly we define $\mu_j^\text{Q}$ as the mean value of the pruned jet mass for these $W$-like interpretations for the same jet.
Thus we have
\begin{equation}
\label{eq:muj}
\mu_j^\text{Q} = \frac{1}{\tau_j N_\text{iter}} \sum_{k \backepsilon\, m_{j,k} \in  \: \Omega} m_{j,k} 
\ = \   \frac{\int_\Omega f_j (m_j) m_j dm_j}{\int_\Omega f_j (m_j) dm_j}\,.
\end{equation}
For comparison, $\mu_a$ in Figure~\ref{fig:window} indicates the average jet mass for the full distribution, not just in the signal window.  For a
background (QCD) jet, this full-average mass value
is generally quite different from $\mu_j^\text{Q}$. 

Let us quickly review the Qjets procedure up to this point.  We begin with a choice of the jet finding algorithm and kinematic cuts, \textit{e.g.}, 
the anti-$k_T$ jet algorithm with $R = 1.0$ and kinematic cuts on the jet, $p_T > 200\gev$ and rapidity $|y| \leq 1.0$.  Then we subject the jets 
identified in this fashion to the Qjets procedure with specific choices of the Qjets parameters $\alpha$ and $N_\text{iter}$ to produce the single-jet 
pruned mass distribution in  Figure~\ref{fig:window}.  With a specific signal jet in mind, say boosted W-jets, we define the mass window $\Omega$ in  Figure~\ref{fig:window}.  This procedure results in values for $\tau_j^\text{Q}$ and $\mu_j^\text{Q}$ from Eqs.~\eqref{eq:tau} and ~\eqref{eq:muj}, which provide a measure of the likelihood that the given jet is a signal jet along with an estimate of the ``true'' mass of that signal jet.

\section{\label{sec:pheno}Results from phenomenological studies}

As an introduction to the following discussion of the results of our phenomenological studies, recall that the goal of the current work is to provide a more detailed explanation of the claim 
made in Ref.~\cite{Ellis:2012sn} that the Qjets procedure improves the statistical stability of jet observables. The fundamental point is that, unlike a conventional binary tagging algorithm that identifies a jet as either tagged or not, the Qjets procedure yields a continuously valued tagging probability 
for a jet (as detailed in  Section~\ref{sec:review}). If observables are constructed 
using these tagging probabilities, it is non-trivial to estimate the statistical uncertainties associated with these observables as the tagging 
probabilities exhibits 
a continuous distribution on the 
interval $\left[ 0,1 \right]$. For example, the well known result that the statistical uncertainty associated with the measurement of the number of 
tagged jets is given by 
$\delta N_T = \sqrt {N_T}$, is no longer true. In Section~\ref{sec:statuncer}, we gave analytic expressions for these uncertainties. In this section, we report the results of our phenomenological studies, where we analyze carefully prepared event samples (generated by standard Monte Carlo event generators) and use the formulas from 
Section~\ref{sec:statuncer} to demonstrate that indeed the Qjets procedure improves the
statistical uncertainties associated with cross-section and mass measurements. 

To be specific, we study the problem of tagging jets containing the decay products of $W$-particles.  We treat a  set of $WW$ diboson events, where both $W$s decay hadronically, 
as signal events. We also consider QCD dijet events that provide the primary background to $W$-tagging. We generate both signal and background events for a $14\tev$ LHC, using Pythia~$8$~\cite{Sjostrand:2007gs}. Additionally, we use the ``ATLAS UE Tune AU2-CTEQ6L1''~\cite{ATLAS:2011zja}  provided by Pythia~$8$ to give these events a realistically 
busy environment corresponding to 
actual proton-proton collisions. The detector simulation is provided by Delphes~\cite{deFavereau:2013fsa}. In particular, we use the default parameters provided by Delphes to simulate the ATLAS detector. Delphes output consists of energy flow four-vectors that are constructed out of the calorimeter cells, tracks, and muon elements of the detector. We do not impose any additional cut on rapidity or $p_T$ on the Delphes output. We cluster the Delphes outputs into anti-$k_T$ jets with $R= 0.7$  and $p_T > 500\gev$ using Fastjet~\cite{Cacciari:2011ma}.  Only the leading jet from each event is selected for further analysis.  

We perform the Qjets procedure using the publicly available Qjets plugin\footnote{http://jets.physics.harvard.edu/Qjets}. The constituents of the selected jets are reclustered for various values of the rigidity parameter (listed in Table~\ref{table:stat}) using the C/A definition of the separation metric (see Eq.~\eqref{eq:dij_defns}).  For the pruning parameter $D_\text{cut}$ we use $D_\text{cut} = m/p_T$, where $m$ and $p_T$ are the mass and the transverse momentum of the unpruned jet respectively.  We perform our analysis for two $z_\text{cut}$ parameter values, $0.1$ and $0.15$, where the smaller value corresponds to the default or optimized pruning case and the larger value should lead to a bit of ``over''-pruning.  Results for both of these values are listed in   Table~\ref{table:stat}.   Finally, we set the $W$-mass window to $\left(70-90\right)\gev$ for the purpose of tagging. 

At this point, we reiterate that in this work we are interested in \textit{both} the effects of a switching from a binary to a continuous tagging variable and from the 
corresponding change in the weighted  average mass.  In order to define a separation of these effects we introduced a new binary tagging efficiency $\tilde{\tau}^{{\text{Q}} }$ 
for every $\tau^\text{Q}$ obtained after the Qjets procedure (as defined in the Introduction). 
The ratio of the statistical uncertainty estimated using $\tau^\text{Q}$ to that using $\tilde{\tau}^{{\text{Q}} }$ therefore provides an estimate of 
the statistical improvement arising primarily from the differences between binary and continuous tagging variables (with the identical mass distribution).  
We also consider the differences between an analysis using $\left( \mu^\text{Q},\tilde{\tau}^{{\text{Q}} }\right)$ versus one using  
$\left(\mu^\text{C},{\tau}^{{\text{C}} }\right)$ to try to isolate the effects primarily due to the changes in the mass distribution 
(which we label ``physics'' effects).

To introduce our explicit numerical results it will be useful to make a few more comments to define the notation used:
\begin{itemize}

\item As noted above, we are studying both a sample of $W$-jets, or signal jets, and QCD-jets, or background jets.  The corresponding results will be labeled by $S$ and $B$.

\item We also include results for the hybrid analysis of Eq.~\eqref{eq:tilde} that is intended to separate statistical from physics effects.  
In particular, since this analysis 
uses a binary $\tilde \tau^\text{Q}$ (with values only $0$ and $1$), the corresponding average tagging efficiency and fluctuation are given by 
$\langle \tilde{\tau}^\text{Q}\rangle =  \tilde{\epsilon}$ and $ \sigma_{\tilde{\tau}}^\text{Q} = \tilde{\epsilon} \left( 1 - \tilde{\epsilon}\right)$ 
respectively (see Eq.~\eqref{eq:momentstilde}). 
The uncertainties associated with the measurement of the cross-section and mass in this hybrid analysis 
can be estimated from the corresponding formulas for conventional analysis in Eq.~\eqref{eq:dnnrconv} and Eq.~\eqref{eq:mrvarNconv}, respectively, 
using the substitutions $\epsilon \rightarrow \tilde{\epsilon} $, 
$\langle \mu^\text{C} \rangle \rightarrow \langle \tilde{\mu}^\text{Q} \rangle $, and 
$\sigma^\text{C}_\mu \rightarrow \sigma_{\tilde{\mu}}^\text{Q} $.
Once again we follow the convention that the appearance of $\tilde{\tau}$ and $\tilde{\mu}$ in these moments reflects the fact that these moments are calculated from their definitions in Eqs.~(\ref{eq:moments}, \ref{eq:moments2}, \ref{eq:omegamoments}) using the hybrid pdf $\widetilde{F}_1^\text{Q}$,
which we discuss in more detail below.

\end{itemize}

The statistical quantities we look at are given by the following equations:
\begin{equation}
\label{eq:finalexpr}
\begin{split}
&  \frac{\delta S^\text{Q}/\sqrt{S^\text{Q} } } {\delta \tilde{S}^{{\text{Q}} } / \sqrt{\tilde{S}^{{\text{Q}} }  } } =
	\frac{\delta S^\text{Q}}{\sqrt{S^\text{Q} }} = 
		\sqrt{\langle \tau_S \rangle + \frac{\sigma_{\tau_S}^2 }{ \langle \tau_S \rangle }  } \, , \qquad 
\frac{\delta B^\text{Q}/\sqrt{B^\text{Q} } } {\delta \tilde{B}^{{\text{Q}} } / \sqrt{\tilde{B}^{\widetilde{\text{Q}} }  } } =
	\frac{\delta B^\text{Q}}{\sqrt{B^\text{Q} }} = 
		\sqrt{\langle \tau_B \rangle + \frac{\sigma_{\tau_B}^2 }{ \langle \tau_B \rangle }  }
                              \\ &
\frac{S^\text{Q}/\delta B^\text{Q}  }{\tilde{S}^{{\text{Q}} }/ \delta \tilde{B}^{{\text{Q}} } } =
	\left(\frac{\langle \tau_S \rangle }{\tilde{\epsilon}_S } \right)   \times
		\sqrt{\frac{\tilde{\epsilon}_B } { \langle \tau_B \rangle } }  \times 
	\frac{1}{\sqrt{ \langle \tau_B \rangle + \frac{\sigma_{\tau_B}^2 }{ \langle \tau_B \rangle } }   }\, , \\  &
\frac{\delta m_T^\text{Q}/m_T^\text{Q}}{\delta \tilde{m}_T^{{\text{Q}} }/\tilde{m}_T^{{\text{Q}} }	} = 	
	\sqrt{ \left( \frac{\sigma_{\mu_S \tau_S}^2}{\langle \mu_S \tau_S \rangle^2} 
		+ \frac{\sigma_{\tau_S}^2}{\langle \tau_S \rangle^2}   
			- 2  \frac{\sigma(\tau_S, \mu_S \tau_S) }{\langle \tau_S \rangle \langle \mu_S\tau_S \rangle}
					 \right) / 
			\left( \frac{\sigma_{\tilde{\mu}_S}^2}{\tilde{\epsilon}_S \langle \tilde{\mu}_S \rangle^2} 
				  \right)  }	 \; ,
\end{split}
\end{equation}
where we have used the equations derived in Section~\ref{sec:statuncer}.

\begin{table}
\centering
	\begin{tabular}{| c ||  c | c  || c | c || c | c || c | c || c | c || c | c |  }
		\hline
		& \multicolumn{8}{c||} {Statistical Effects} & \multicolumn{4}{c|} {Total uncertainty }  \\
		\cline{2-13}
		
		\multirow{4}{*}{$\alpha$ } 
		& \multicolumn{2}{c||}{\multirow{2}{*} { $\frac{\delta S^\text{Q} }{ \sqrt{S^{\text{Q} } } }$}  }
		& \multicolumn{2}{c||}{\multirow{2}{*}  {$\frac{\delta B^\text{Q} }{ \sqrt{B^{\text{Q} } } }$}  }
		& \multicolumn{2}{c||} {\multirow{2}{*} {$\frac{S^\text{Q} / \delta B^\text{Q} } {\tilde{S}^{\text{Q}} /
				 \delta \tilde{B}^{\text{Q}} }$}  } 
		& \multicolumn{2}{c||} {\multirow{2}{*} {$\frac{\delta m_T^\text{Q}/m_T^\text{Q}}
				{\delta \widetilde{m}_T^{\text{Q} }/\widetilde{m}_T^{\text{Q} }	} $}  }         
		& \multicolumn{2}{c||} {\multirow{2}{*} {$\frac{S^\text{Q} / \delta B^\text{Q} }
				 {S^{\text{C}} /\delta B^{\text{C}} }$}  } 
		& \multicolumn{2}{c|} {\multirow{2}{*} {$\frac{\delta m_T^\text{Q}/m_T^\text{Q}}
				{\delta m_T^{\text{C}} /m_T^{\text{C}}	} $}  }         \\
					
		& \multicolumn{2}{c||}{ }  & \multicolumn{2}{c||}{ } &
				\multicolumn{2}{c||}{ }   & \multicolumn{2}{c||}{ }&
				\multicolumn{2}{c||}{ }   & \multicolumn{2}{c|}{ }  \\
				
		& 	\multicolumn{2}{c||} { $z_\text{cut}$ } & \multicolumn{2}{c||} { $z_\text{cut}$ } & 
				\multicolumn{2}{c||} { $z_\text{cut}$ } &\multicolumn{2}{c||} { $z_\text{cut}$ }& 
					\multicolumn{2}{c||} { $z_\text{cut}$ } &\multicolumn{2}{c|} { $z_\text{cut}$ } \\	
				
			 &  {$ 0.10$}  & {$ 0.15$} 			 
			 &  {$ 0.10$}  & {$ 0.15$}
			 &  {$ 0.10$}  & {$ 0.15$} 
			 &  {$ 0.10$}  & {$ 0.15$ } 
			 &  {$ 0.10$}  & {$ 0.15$} 
			 &  {$ 0.10$}  & {$ 0.15$ }\\
		\hline
		\hline
	\ 	10.0	& 0.99	& 0.98	& 0.94	& 0.94	& 1.17	& 1.15	& 1.00	& 1.01  	
				& 1.05	& 1.05	& 0.96	& 0.96  \\	
	\ 	1.00 	& 0.95	& 0.94 	& 0.85	& 0.85	& 1.42	& 1.38	& 1.00	& 1.05 	
				& 1.16	& 1.17	& 0.86	& 0.86  \\
	\ 	0.10 	& 0.90	& 0.88	& 0.74	& 0.72	& 1.63	& 1.57	& 1.00	& 1.08 	
				& 1.26	& 1.29	& 0.73	& 0.71  \\	
	\	0.01 & 0.86	& 0.82	& 0.69	& 0.66	& 1.61	& 1.54	& 0.98	& 1.00	
				& 1.22	& 1.25	& 0.65	& 0.56  \\
	\	0.00	& 0.87	& 0.82	& 0.77	& 0.72	& 1.28	& 1.24	& 0.88	& 0.92	
				& 1.00	& 1.03	& 0.60	& 0.52  \\
		\hline
	\end{tabular}					
\caption{\label{table:stat}  Statistical uncertainties associated with various measurements of cross-section and mass. Formulas used to estimate these quantities as listed in Eq.~\eqref{eq:finalexpr}.}
\end{table}

In Table~\ref{table:stat}, we tabulate the numerical estimations of the various observables for different values of $z_\text{cut}$ and $\alpha$.  In 
the remaining part of this section we provide a brief description of the patterns observed in Table~\ref{table:stat}. Detailed explanations of these 
observations will be provided in the following two sections.

The first four observables in the table capture what we have labeled 
the statistical improvements seen in the Qjets procedure for the signal and  the background samples. 
The quantity $\delta N_T^{\text{Q}} / \sqrt{{N}_T^{\text{Q}}}$ for both signal and background  represents the improvement 
in the uncertainty of the measured cross-section due to what we have labeled statistical effects. 
Note that this quantity is unity for a binary tagging variable, which is why the denominator becomes unity in the first line of Eq.~\eqref{eq:finalexpr}. 
For large $\alpha$ these quantities are close to $1$ for both
values of the $z_\text{cut}$ parameter. This situation reflects the fact that at high rigidity the individual mass distribution for each jet is quite narrow
even after applying the Qjets procedure
and $\tau$  mostly has the values $0$ or $1$ (\textit{i.e.}, the Qjets procedure approaches  
the ``classical'' limit as $\alpha \rightarrow \infty$).  As indicated in  Table~\ref{table:stat},  the uncertainties decrease as $\alpha$ is 
decreased and we include an increasing range of different clustering/pruning scenarios until a plateau is reached at $\alpha \sim 0.01$ 
(for the background the uncertainty actually turns over and starts to
increase again as $\alpha \rightarrow 0$).  It is interesting also to note that the improvement with decreasing $\alpha$ is slightly better (\textit{i.e.},
smaller values of the ratio) for the less optimal $z_\text{cut}$ value (0.15).  This feature presumably arises from the fact that we start, in the classical
limit, with less than optimal pruning, which allows the Qjets procedure more opportunity to include different clustering/pruning scenarios that improve 
the situation.  The background case is somewhat less $z_\text{cut}$ dependent as expected, as there is less of a clear definition of optimal pruning.

The statistical improvement in the discovery potential is captured by the third quantity, the ratio 
$\left( S^\text{Q} / \delta B^\text{Q} \right) /\left( \tilde{S}^\text{Q} /\delta \tilde{B}^\text{Q} \right)$. The larger this number becomes, the better 
is the chance that a precise measurement of the signal can be performed with a given luminosity. Once again we see that 
this observable is  maximized for a small $\alpha \sim 0.01$.  The small $z_\text{cut}$ dependence in this case makes the not unexpected suggestion
that it is best for the Qjets procedure to perturb around an optimal classical choice of parameters.   

Finally, the fourth observable in Table~\ref{table:stat} provides an estimate of the uncertainty associated with the measurement of the jet mass arising from what we have labeled as statistical effects.  We interpret the fact that this ratio
remains near unity (except for very small values of $\alpha \sim 0$) as confirmation that we have largely succeeded in separating
the effects of binary versus continuous tagging variables, which we see are small for this variable, from the effects of changing the mass
distribution itself, which will be important for this quantity.   
We refer the reader to Section~\ref{sec:understanding} for further explanation of these observations.

For completeness we include our estimate of the total improvements provided by the Qjets procedure using the last two observables in Table~\ref{table:stat}.  These observables compare the uncertainties in the Qjets procedure to those in the conventional or classical procedure. As explained earlier, these quantities can be calculated from Eq.~\eqref{eq:finalexpr} by the replacements $\tilde{\epsilon} \rightarrow \epsilon$,  $ \langle \tilde{\mu}^\text{Q} \rangle \rightarrow \langle \mu^\text{C} \rangle$, and  $\sigma_{\tilde{\mu}}^\text{Q} \rightarrow \sigma^\text{C}_\mu $.  
Overall we find that the behavior of the  statistical uncertainties associated with the cross-section and mass is similar to what was described in Ref.~\cite{Ellis:2012sn}. The cross-section measurement is most stable in the range  $0.1 \geq  \alpha >  0.01$, whereas the mass uncertainty prefers even smaller rigidity ($0.01 >  \alpha \geq  0.0$).

Note that the contribution to the uncertainties from what we have labeled physics effects can be found by simply dividing the total uncertainty by the corresponding 
statistical contribution. These results will be discussed in more detail in  Section~\ref{sec:expl}.  It is worthwhile noting that this exercise already tells
us that the effects we labeled physics will be more important than the statistical effects for the mass measurement uncertainties, as we just suggested.

 \begin{figure}[!ht]
	\centering 
	{\includegraphics[width=0.7\textwidth]{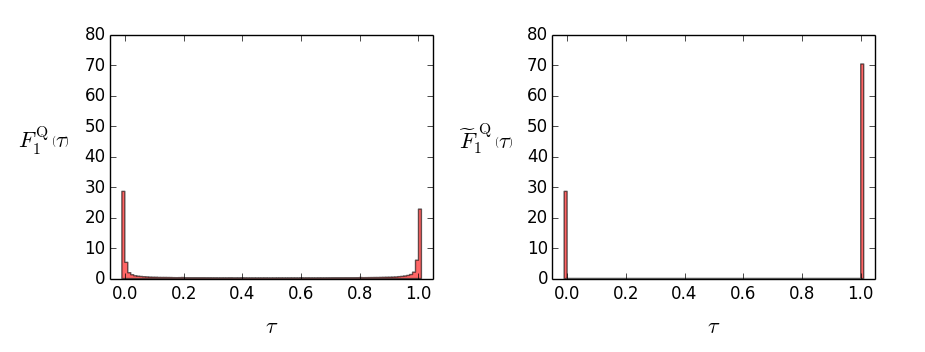}  \label{fig:WF1QtA10}} \\
	{\includegraphics[width=0.7\textwidth]{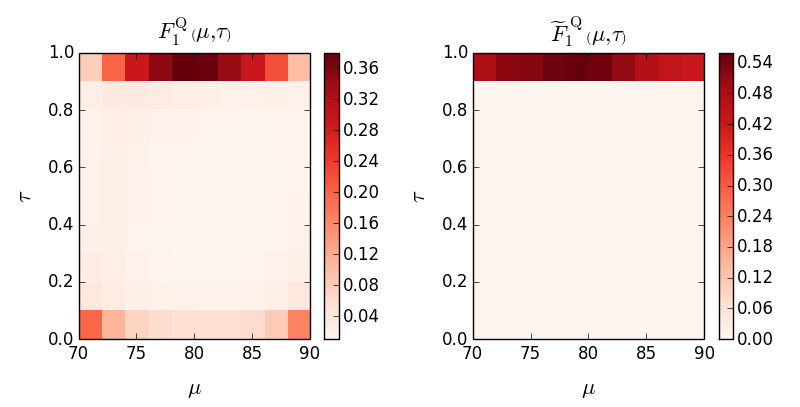} \label{fig:WF1QmtA10} }
\caption{The probability distributions $F_1^\text{Q} (\tau), \widetilde{F}_1^{{Q}} (\tau), F_1^\text{Q} (\mu,\tau), \widetilde{F}_1^{{Q} }(\mu,\tau)$ derived from a sample of $W$-jets. These particular distribution are produced using $\alpha= 1.0$ and  $z_\text{cut} = 0.1$. For the rest of the parameters, see  Section.~\ref{sec:pheno}.}
\label{fig:F1QA10}
\end{figure}

%------------------------------------------------------------------------------------------------------------------
\section{\label{sec:understanding}Understanding the statistical effects}

In order to understand the uncertainties listed in Table~\ref{table:stat} it is essential to study the probability distributions $F_1 (\tau)$ and $F_1(\mu,\tau)$. 
In Figure~\ref{fig:F1QA10}, we display these distributions as derived from a sample of $W$-jets. On the left are the distributions 
$F_1^\text{Q} (\tau)$ and $F_1^\text{Q}(\mu,\tau)$
arising from the full Qjets analysis, while the plots on the right illustrate the distributions $\widetilde{F}_1^\text{Q} (\tau)$ 
and $\widetilde{F}_1^\text{Q}(\mu,\tau)$ from the hybrid analysis. 
Recall that the latter analysis uses the binary tagging probability $\tilde{\tau}^\text{Q}$ derived from the standard Qjets 
probability $\tau^\text{Q}$ as defined in Eq.~\eqref{eq:tilde}, \textit{i.e.}, all nonzero $\tau^\text{Q}$ values ($\tau^\text{Q} > 0$) correspond to
$\tilde{\tau}^\text{Q} = 1$.
By construction, $F_1^\text{Q} (\tau=0) = \widetilde{F}_1^{\text{Q}} (\tau=0)$ as illustrated by the equal heights of the zero bins of 
$F_1^\text{Q}(\tau)$ and $\widetilde{F}_1^{\text{Q}} (\tau)$ in
Figure~\ref{fig:F1QA10}.  
The difference between the two distributions arises from the fact that the rest of the probability in $\widetilde{F}_1^{\text{Q}} (\tau)$ all lies in 
the $\tau=1$ bin, whereas $F_1^\text{Q}(\tau)$ exhibits nonzero probability at values of $\tau$ between $0$ and $1$ (although it is still strongly peaked
in the $\tau=1$ bin). In other words in moving from the $\widetilde{F}_1^{\text{Q}} (\tau)$ distribution to the $F_1^\text{Q}(\tau)$ distribution (\textit{i.e.}, moving from
a binary tagging probability to a continuous one), probability ``leaks out'' of the $\tau=1$ bin into the $1>\tau>0$ bins.  

The lower plots of $F_1^\text{Q} (\mu,\tau)$
and $\widetilde{F}_1^{\text{Q} }(\mu,\tau)$ provide additional information.  In particular, almost all jets that leak-out of the $\tau=1$ bin, as one moves from 
$\widetilde{F}_1^{\text{Q} }(\mu,\tau)$ to $F_1^\text{Q} (\mu,\tau)$, lie near or at one of the boundaries of the window in $\mu$.  Also note that the distribution in $\mu$ corresponding to $\tau = 1$ is peaked near the $W$ mass (as expected for an underlying $W$-jet sample), and that the $\tau = 0$ bin does not actually  appear in the lower plots as all
of the corresponding $\mu$ values are \textit{outside} of the $\mu$ window (by definition).   Lastly, but perhaps most importantly, if we sum over $\tau$ but with no explicit $\tau$
weighting, the resulting mass distributions are identical, $\int d\tau \widetilde{F}_1^{\text{Q} }(\mu,\tau) = \int d\tau F_1^\text{Q} (\mu,\tau)$.
To make this last point explicit we note the following results for the moments of these two distributions,
\begin{eqnarray}
 \tilde{N}_\Omega^\text{Q} \ &=& \   \int_\Omega d\mu \int_0^1 d\tau\,  \tilde{F}_1^\text{Q}(\mu,\tau) 
                      \ = \  \tilde{\epsilon} \ = \ \langle \tilde{\tau}^\text{Q} \rangle  \ = \ 
		\int_\Omega d\mu \int_0^1 d\tau  \,  F_1^\text{Q}(\mu,\tau)    \ = \ 
		{N}_\Omega^\text{Q} \,, \notag \\
\langle \tilde{\mu}^\text{Q} \rangle \ &=& \ \frac{1}{\tilde{N}_\Omega^\text{Q}}  \int_\Omega d\mu \int_0^1 d\tau\, \mu \tilde{F}_1^\text{Q}(\mu,\tau) \ = \ 
		\frac{1}{{N}_\Omega^\text{Q}}  \int_\Omega d\mu \int_0^1 d\tau\, \mu {F}_1^\text{Q}(\mu,\tau) = \langle \mu^\text{Q} \rangle \,, \notag \\
\left(\sigma_{\tilde{\mu}}^\text{Q}\right)^2  \ &=& \ \frac{1}{\tilde{N}_\Omega^\text{Q}}  \int_\Omega d\mu \int_0^1 d\tau\, 
(\mu - \langle \tilde{\mu}^\text{Q} \rangle)^2
         \tilde{F}_1^\text{Q}(\mu,\tau) \ = \ 
		\frac{1}{{N}_\Omega^\text{Q}}  \int_\Omega d\mu \int_0^1 d\tau\,  (\mu - \langle \mu^\text{Q} \rangle)^2 {F}_1^\text{Q}(\mu,\tau) \notag \\
                \ &=& \ \left(\sigma_{{\mu}}^\text{Q}\right)^2  \,.
\label{eq:results5}
\end{eqnarray}
These equalities should help to confirm that comparing the Qjet and $\widetilde{\text{Q}}$jet analyses, as in Table~\ref{table:stat}, 
focuses on the statistical effects, while
comparing the $\widetilde{\text{Q}}$jet analysis with the conventional analysis focuses primarily on the physics effects caused by the changes in the mass distributions, as we will discuss in Section~\ref{sec:expl}.

With these insights, we can construct an explicit toy model that helps to illuminate the connection between the two distributions $F_1^\text{Q}$ and $\widetilde{F}_1^{\text{Q}} $. 
We can approximate the filled bins (closest to the boundaries) as being described by delta functions (recall the description of the conventional result in Eq.~\eqref{eq:convF1}).
Considering first adding just a single extra bin near the upper boundary, we have
\begin{equation}
\begin{split}
\label{eq:toyF1t}
	\widetilde{F}_1^{\text{Q}}\left( \tau \right) &\ = \ (1 - \tilde{\epsilon} )\ \delta (\tau) \ + \ \tilde{\epsilon} \	
			\delta(\tau-1)\,  \\  
	F_1^\text{Q} \left(\tau \right) &\ \simeq \ \widetilde{F}_1^{\text{Q}} \left( \tau \right)  - \Delta \big[ \delta \left( 1-\tau \right) -
			\delta\left( 1-\eta - \tau \right) \big]   \, ,
\end{split}
\end{equation}
where (as shown in Eq.~\eqref{eq:tildeF1}) $\widetilde{F}_1^{\text{Q}}\left(\tau \right) $ is represented by a binomial representation with mean $\tilde{\epsilon} $ and variance $ \sigma^2_{\tilde{\tau}} = \tilde{\epsilon}  \left( 1- \tilde{\epsilon}  \right)$.  The extra term in the expression for $F_1^\text{Q}\left(\tau \right)$
is intended to present the fact that a small fraction of the jets, $\Delta$, have migrated from the $\tau=1$ bin to the $\tau = (1- \eta)$ bin ($ 0 <\eta < 1$). 
It is straightforward to evaluate the corresponding approximate mean and variance of $F_1^\text{Q}\left(\tau \right)$ in the limit $\Delta \ll \tilde{\epsilon} $ 
in terms of the mean and 
variance of $\widetilde{F}_1^{\text{Q}}\left(\tau \right) $.  To first order in $\Delta/ \tilde{\epsilon}$ we find
\begin{equation}
\begin{split}
\label{eq:toyF1tmoments}
\langle \tau \rangle \ &\ \simeq \  \tilde{\epsilon}  - {\Delta } \eta \,,  \\ 
\sigma_\tau^2 &\ \simeq  \tilde{\epsilon}  \left(1- \tilde{\epsilon}  \right)
					+ \Delta  \eta \big( \eta 
					- 2\left(1- \tilde{\epsilon}  \right) \big)
                           = \sigma^2_{\tilde{\tau }} 
					+ \Delta  \eta \big( \eta 
					- 2\left(1-\tilde{\epsilon}  \right) \big)  \; .
\end{split}
\end{equation}
Applying this result to the first few column of Table~\ref{table:stat} we obtain 
\begin{equation}
\label{eq:toyNuncer}
 \frac{\delta S^\text{Q}/\sqrt{S^\text{Q} } } {\delta S^{\widetilde{Q} } / \sqrt{S^{\widetilde{Q} }  } } \ = \
	\frac{\delta S^\text{Q}}{\sqrt{S^\text{Q} }} = 
		\sqrt{\langle \tau_s \rangle + \frac{\sigma_{\tau_s}^2 }{ \langle \tau_s \rangle }  } \ \simeq \
		 1 - \frac{\Delta }{2 \tilde{\epsilon} } \eta  \left( 1- \eta \right)    
		 \ \leq 1 \ \; .
\end{equation}
Noting that this expression is symmetric in $\eta \to 1-\eta$, we see that the bins at both ends of the $\tau$ distribution will contribute in a 
similar fashion, decreasing the
scaled fluctuations in this observable.   So, if we define a more accurate approximate expression for $F_1 (\tau)$, including all of the filled in bins 
($\eta_k$ near $0$ and 
near $1$),
\begin{equation}
F_1^\text{Q}\left(\tau \right) \ \simeq \ \widetilde{F}_1^{\text{Q}} \left( \tau \right)  - \sum_k \Delta_k \big[ \delta \left( 1-\tau \right) -
			\delta\left( 1-\eta_k - \tau \right) \big]   \, ,
\end{equation}
we find 
\begin{equation}
\label{eq:toyNuncer2}
 \frac{\delta S^\text{Q}/\sqrt{S^\text{Q} } } {\delta S^{\widetilde{Q} } / \sqrt{S^{\widetilde{Q} }  } } \ \simeq \
		 1 - \sum_k \frac{\Delta_k }{2 \tilde{\epsilon} } \eta_k  \left( 1- \eta_k \right)    
		 \ \leq 1 \ \; .
\end{equation}
Since $0 \leq \eta_k, \tilde{\epsilon}  \leq 1$ and $\Delta_k > 0$, \textit{all} of the terms in the sum serve to decrease these fluctuations 
(at least to leading order in $\Delta_k/ \tilde{\epsilon} $).  As rigidity is decreased and the analysis moves further from the ``classical'' limit,
we expect more bins away from the edges to be filled-in, which explains, at least qualitatively, the systematic decrease with decreasing rigidity 
(at least until we reach zero rigidity)
in the first two columns, both signal and background, in Table~\ref{table:stat}. 
Note that the deviation of the LHS from $1$ in Eq.~\eqref{eq:toyNuncer2} is essentially proportional to the factor $\sum_k \Delta_k/\tilde{\epsilon} $. 
In our toy example, this represents the fraction of jets that occupied the $\tau = 1$ bin in $\widetilde{F}_1^{\text{Q}}\left(\tau \right)$, but correspond to 
a smaller $\tau$ value in $F_1^\text{Q}\left(\tau \right)$.  As we mentioned earlier, these jets have masses near the boundary of the mass
window and we can say that $\sum_k \Delta_k/\tilde{\epsilon} $ represents the fraction of jets that reside near the mass boundary 
(at least in our toy example) and that, after the full Qjets procedure, exhibit less than unit tagging probability.  

\begin{figure}[h]\centering
	\includegraphics[width=.9\textwidth]{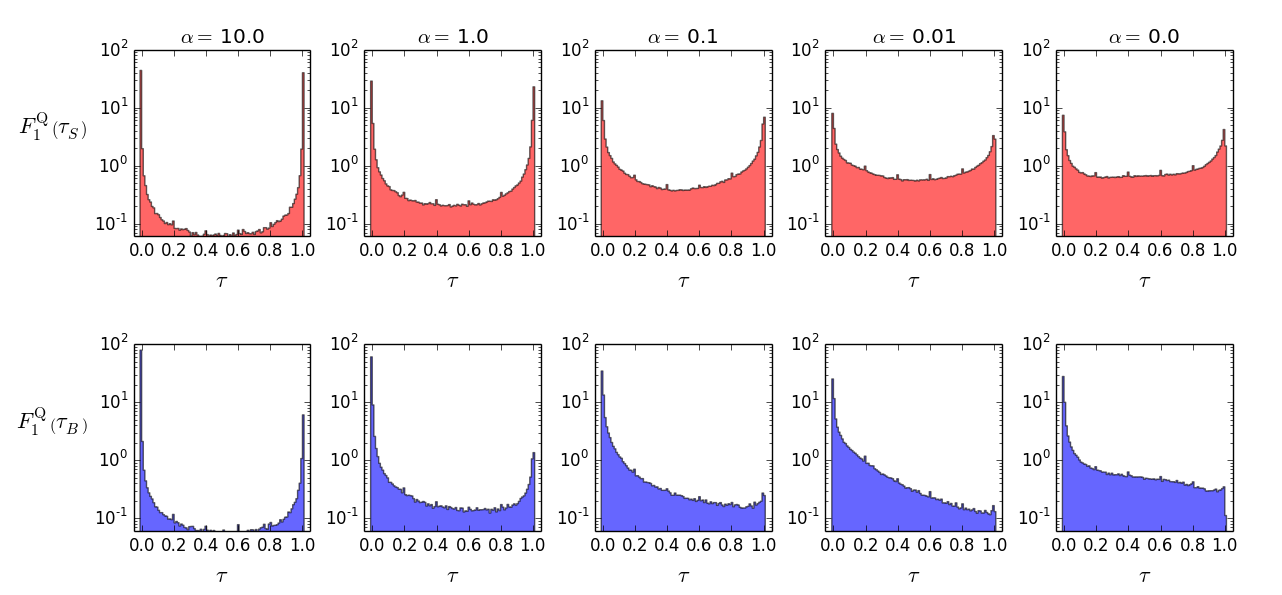} \\
		\includegraphics[width=0.9\textwidth]{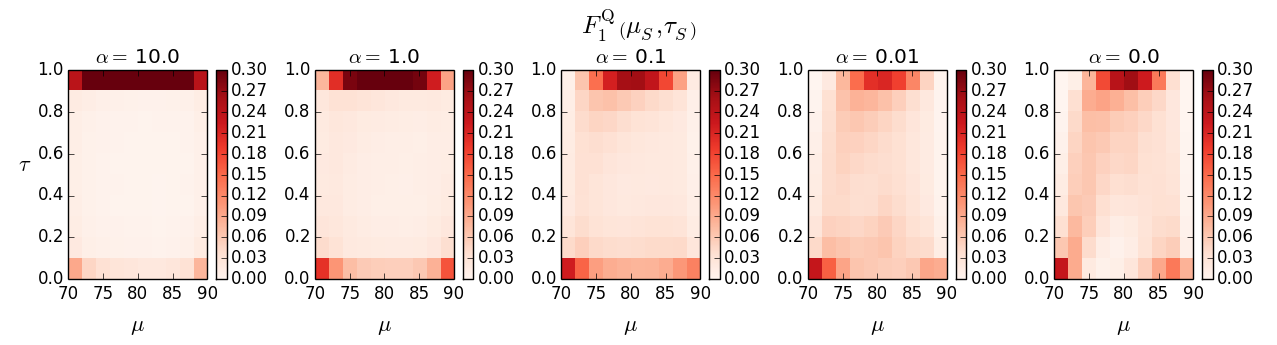} \\
	\includegraphics[width=0.9\textwidth]{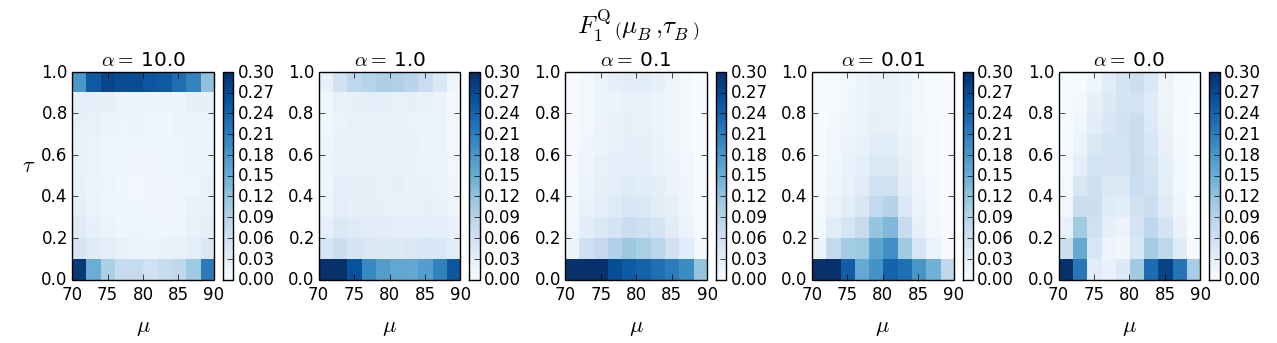}
	\caption{\label{fig:pdfs} The distributions  $F_1(\tau)$ and $F_1(\mu,\tau)$ for signal (in red) and background (in blue) jets as functions of $\alpha$.}
\end{figure}
To provide a more detailed picture of the rigidity ($\alpha$) dependence exhibited in Table~\ref{table:stat},  Figure~\ref{fig:pdfs} shows plots  of
$F_1^\text{Q}\left( \tau \right)$ and $F_1^\text{Q}\left(\mu,\tau \right)$ for various values of the rigidity with $z_\text{cut} = 0.1$.  To appreciate
these plots it is important to note a couple of relevant features.  In order to make visible the values at intermediate $\tau$ values, the plots of $F_1^\text{Q}\left( \tau \right)$ are semi-log plots, while the 2-D $F_1^\text{Q}\left(\mu,\tau \right)$ plots use a linear scale for the color scale.
Further, the $\tau = 0$ bin at the extreme left of the $F_1^\text{Q}\left( \tau \right)$ plots corresponds to $\mu$ values not shown in the 
2-D $F_1^\text{Q}\left(\mu,\tau \right)$ plots, which show only the $\mu$ values in the window $\Omega$. 
To understand the general structure of the plots in Figure~\ref{fig:pdfs} 
recall that, as we lower the value of the rigidity parameter $\alpha$, the Qjets procedure explores an ever broader spectrum of clustering histories. This leads to 
changes in the plots arising from two distinct underlying effects, where both are associated with $\mu$ values near the boundaries of the window $\Omega$.   
Individual jets, whose Qjet mass distributions are entirely within $\Omega$ for large $\alpha$ values, \textit{i.e.}, appear in the
$\tau = 1$ bin, may eventually exhibit Qjets mass distributions with tails that extend
outside of $\Omega$ for sufficiently small $\alpha$ values.  So, as the value of $\alpha$ decreases, such jets will gradually populate the bins with $\tau < 1$,
but still with $\mu$ near the boundaries, at least initially.  Likewise there are jets whose large $\alpha$ Qjets mass distributions are entirely outside of $\Omega$, 
\textit{i.e.}, appear in the $\tau = 0$ bin,
but then develop tails inside of $\Omega$ for sufficiently small $\alpha$ values.  These jets will gradually populate the bins at small $\tau$ values, always moving
inward from the boundaries in $\mu$.  This simple picture is generally correct for both the signal and background samples as illustrated in Figure~\ref{fig:pdfs}.
As $\alpha$ decreases the distributions in $\tau$, $F_1^\text{Q}\left( \tau \right)$, for both signal and background, gradually fill-in at intermediate
$\tau$ values leading to decreasing values of the fluctuation $\sigma_\tau^\text{Q}$.  In turn this means that $\delta N_T/\sqrt{N_T}$ decreases with
decreasing $\alpha$ values as indicated by the numbers in Table~\ref{table:stat}, and the toy model analysis of Eq.~\eqref{eq:toyNuncer2}.

In the limit $\alpha \to 0$ the $F_1^\text{Q}\left(\mu,\tau \right)$ plots for both signal (red) and background (blue) suggest an ``arc'' structure, \textit{i.e.},
a ridge of enhanced probability connecting the $\mu$ boundaries at small $\tau$ via bins at intermediate $\mu$ values at larger $\tau$ values.  The primary distinction between signal and background is the fact that the bins near $\tau = 1$ for the background are rapidly depopulated as $\alpha$ decreases, while for the signal
the bins near $\tau = 1$ maintain a population similar to those near $\tau = 0$ and the bins near $\tau = 1$ always exhibit a $\mu$ distribution with a peak
near the signal mass ($M_W$).

Figure~\ref{fig:pdfs} also indicates that for large rigidity (say, $\alpha \geq 1.0$), the probabilities for finding jets with intermediate $\tau$ values ($0 < \tau < 1$), are tiny. In this range of $\alpha$, the approximations made in our toy model  above are quite appropriate. For smaller rigidity ($\alpha < 1.0$), more and more jets occupy the intermediate $\tau$ values, new patterns emerge in the pdf $F_1^\text{Q}\left(\mu,\tau \right)$, and the deviations of  $F_1^\text{Q}$ from $\tilde{F}_1^\text{Q}$ are not necessarily small.

%------------------------------------------------------------------------------------------------------------------
\section{\label{sec:expl}Understanding the physics effects}

\begin{table}
\centering
\begin{tabular}{| c ||  c | c  || c | c || c | c |  }
\hline
	\multirow{4}{*}{$\alpha$ } 
		&  \multicolumn{2}{c||} {\multirow{2}{*} {$\frac{\tilde{S}^{\text{Q}} /
				 \delta\tilde{B}^{{Q}}  } {S^{\text{C}} /\delta B^{\text{C}} }$}  }
		& \multicolumn{2}{c||}{\multirow{2}{*} { $\widetilde{S}^{\text{Q}} / S^{\text{C} }  $}  }
		& \multicolumn{2}{c|}{\multirow{2}{*}  {$\sqrt{ B^{\text{C} }/\widetilde{B}^{\text{Q}}  }$}  } 
\\
		& \multicolumn{2}{c||}{ } & \multicolumn{2}{c||}{ } & \multicolumn{2}{c||}{ }
\\	
		& 	\multicolumn{2}{c||} { $z_\text{cut}$ } & \multicolumn{2}{c||} { $z_\text{cut}$ }
			& \multicolumn{2}{c||} { $z_\text{cut}$ }
\\
		&  {$ 0.10$}  & {$ 0.15$}         &  {$ 0.10$}  & {$ 0.15$}     &  {$ 0.10$}  & {$ 0.15$}
\\
	\hline
	\hline
	10.0	& 0.90	& 0.91	& 1.16	& 1.18	& 0.776	       & 0.774	\\
	1.00	& 0.82	& 0.85	& 1.48	& 1.61	& 0.554	       & 0.530    \\
	0.10	& 0.77	& 0.82	& 1.81	& 2.12	& 0.427	       & 0.385    \\
	0.01	& 0.76	& 0.81	& 1.91	& 2.32	& 0.399	       & 0.351    \\
	0.00	& 0.78	& 0.83	& 1.93	& 2.37	& 0.406	       & 0.350    \\		
\hline
\end{tabular}					
\caption{\label{table:understanding-physics-cs} The physics component of the cross-section uncertainties as functions of $z_\text{cut}$ and rigidity $\alpha$, found by dividing the total uncertainty by the purely statistical component (from Table.~\ref{table:stat}). We also provide numerical values of different components in Eq.~\eqref{eq:Ctilde} responsible for the physics part of the cross-section uncertainties. }
\end{table}

We list the components of the cross-section and mass uncertainties due to what we have labeled physics effects in Table~\ref{table:understanding-physics-cs} and Table~\ref{table:understanding-physics-mass} respectively. The numerical values of these components can be evaluated by dividing the total uncertainty  in Table~\ref{table:stat} by its statistical part (also listed in  Table~\ref{table:stat}).  

Understanding these physics quantities is relatively easier since one does not need to think of fractional tagging efficiencies. In particular,
since both the conventional 
and hybrid analyses use binary tagging efficiencies, the fluctuations in the number of jets goes like $1/\sqrt{N}$ (recall the discussion in Section \ref{sec:statuncer}).
For example, the quantity in the Table~\ref{table:understanding-physics-cs}, 
which measures the improvement in the cross-section measurement significance, simplifies to 
\begin{equation}
 \frac{\widetilde{S}^{\text{Q} } /\delta \widetilde{B}^{\text{Q}}  } {S^\text{C}  /\delta B^\text{C} }  \ = \ \left( \frac{\widetilde{S}^{\text{Q}}  }
{S^\text{C} } \right) \times 
 		\sqrt{ \frac{ B^\text{C} }{\widetilde{B}^{\text{Q} } }   } \; .
\label{eq:Ctilde}
\end{equation}
The improvement in statistical stability, therefore, depends on two independent ratios, the relative signal efficiency ($\widetilde{S}^{\text{Q} }/S^\text{C}$) and
(1 over) the square root of the relative background efficiency ($\widetilde{B}^{\text{Q} }/B^\text{C}$), for the hybrid Qjets analysis compared to the 
conventional analysis.  Table~\ref{table:understanding-physics-cs} separately exhibits the variation of these two components of Eq.~\eqref{eq:Ctilde}  
with the rigidity parameter $\alpha$.
To understand the exhibited behavior, we must recall our previous discussion.  As we decrease $\alpha$, we include new clustering histories, and,
as a result, find that jets, which were previously not tagged (for larger $\alpha$ values), are now tagged. 
By construction $\tilde{\tau}^\text{Q} =1$ for these jets (even though they may have small $\tau^\text{Q}$).  This is why the ratio $\widetilde{N}_T^{\text{Q}}/N_T^\text{C}$ increases with decreasing $\alpha$ for both signal and background. In the case of the (signal)
$W$-jets, almost all jets are tagged even for large $\alpha$ and so $\widetilde{S}^{\text{Q} }/S^\text{C}$ increases relatively 
slowly (but monotonically) as $\alpha$ decreases.  In the language of the simple model in the previous Section (see Eq.~(\ref{eq:toyNuncer2})), the 
behavior of the ratio $\widetilde{S}^{\text{Q} }/S^\text{C}$ is telling us about the magnitude of the leak-in effect, $\sum_k\Delta_k/\tilde{\epsilon}$, at 
least quantitatively (note that the effect is no longer small as $\alpha$ approaches zero).

In the case of background jets, there are always more untagged jets than tagged ones, some of which can be tagged when we allow a broader range of clustering histories as $\alpha$ decreases.
Thus the ratio $\widetilde{B}^{\text{Q} }/B^\text{C}$ \textit{increases} quite rapidly with decreasing $\alpha$, resulting in the somewhat slower
but still rapid \textit{decrease} of the factor $\sqrt{B^\text{C}/\widetilde{B}^{\text{Q} }}$ .  
By Eq.~\eqref{eq:Ctilde}, the physics component of the cross-section uncertainty in Table~\ref{table:understanding-physics-cs} is the product of  
the corresponding values in the two right-hand columns in the Table.  Numerically the decrease of the background ratio
is dominant, leading to a slowly decreasing cross-section uncertainty in the hybrid Qjets analysis compared to the conventional
analysis
with decreasing $\alpha$ until $\alpha$ reaches $0.01$.  
For even smaller $\alpha$ values the sampling of clustering histories is so broad that the qualitative behavior of the background ratio changes
and the relative fluctuations begin to grow.  

Note also that the variation with $\alpha$ of the individual ratios, and the product, is somewhat stronger 
for the non-optimal $z_\text{cut}$ value (0.15).  This is to be expected as the non-optimal conventional result implies that more of the added 
clustering histories in the Qjets analysis will correspond to an improvement.  Note that this is a statement about the \textit{improvement} in the statistical
stability.  Overall one is better off starting with an optimal choice of the conventional pruning parameters to perform the Qjets procedure around.  However,
the results in Table~\ref{table:understanding-physics-cs} do suggest that the Qjets procedure can help to moderate the impact of any initial poor choice
of parameters. 

\begin{table}
\centering
\begin{tabular}{| c ||  c | c  || c | c || c | c || c | c |  }
\hline
	\multirow{4}{*}{$\alpha$ } 
		&   \multicolumn{2}{c||}{\multirow{2}{*}  
			{ $\frac{\delta \widetilde{m}_T^{\text{Q} }/\widetilde{m}_T^{\text{Q}} }
				{\delta m_T^{\text{C}} /m_T^{\text{C}}	} $ }}
		&   \multicolumn{2}{c||}{\multirow{2}{*} {$\sqrt{ S^\text{C} /\widetilde{S}^{\text{Q} } }  $} }
		&   \multicolumn{2}{c||}{\multirow{2}{*} {$\sigma_{\tilde{\mu}_S}^\text{Q}/\sigma_{\mu_S}^\text{C} $} }
		&   \multicolumn{2}{c|}{\multirow{2}{*} {$\langle \mu_S^\text{C} \rangle / \langle \tilde{\mu}_S^\text{Q} \rangle$} }
\\
		& \multicolumn{2}{c||}{ } & \multicolumn{2}{c||}{ } & \multicolumn{2}{c||}{ } & \multicolumn{2}{c|}{ }
\\
		& 	\multicolumn{2}{c||} { $z_\text{cut}$ } & 	\multicolumn{2}{c||} { $z_\text{cut}$ } 
			& 	\multicolumn{2}{c||} { $z_\text{cut}$ }  & 	\multicolumn{2}{c|} { $z_\text{cut}$ } 
\\
		&  {$ 0.10$}  & {$ 0.15$}  &  {$ 0.10$}  & {$ 0.15$}  &  {$ 0.10$}  & {$ 0.15$}  &  {$ 0.10$}  & {$ 0.15$}  
\\		\hline
	10.0	& 0.96	& 0.95	& 0.93	& 0.92	& 1.03	& 1.03	& 1.00	& 1.00  \\
	1.00 & 0.86	& 0.82	& 0.82	& 0.79	& 1.05	& 1.04	& 1.00	& 1.00  \\
	0.10 & 0.73	& 0.66	& 0.74	& 0.69	& 0.98	& 0.95	& 1.01	& 1.01  \\
	0.01 & 0.66	& 0.56	& 0.72	& 0.66	& 0.91	& 0.86	& 1.01	& 1.02  \\
	0.00	& 0.69	& 0.57	& 0.72	& 0.65	& 0.95	& 0.86	& 1.01	& 1.02  \\
\hline
\end{tabular}					
\caption{\label{table:understanding-physics-mass}  The physics component of the mass uncertainties as functions of $z_\text{cut}$ and rigidity $\alpha$, found by dividing the total uncertainty by the purely statistical component (from Table.~\ref{table:stat}). We also provide numerical values of different components in Eq.~\eqref{eq:Ctilde2} responsible for the physics part of the mass uncertainties. }
\end{table}

Finally we turn to the mass measurement uncertainties as described by the results in Table~\ref{table:understanding-physics-mass}. The general expressions from Eqs.~\eqref{eq:mrvarN} and ~\eqref{eq:mrvarNconv} for the signal sample yield 
\begin{equation}
\frac{\delta \widetilde{m}_T^{\text{Q}}/\widetilde{m}_T^{\text{Q}}}{\delta m_T^\text{C}/m_T^\text{C }	} \ = \	
	\sqrt{\frac{\langle \tau^\text{C}_S\rangle}{\langle\tilde{\tau}_S^\text{Q}\rangle}} \times
		\frac{\sigma^{\text{Q}}_{\tilde{\mu}_S}} {\sigma_{\mu_S}^\text{C}} \times
			\frac{\langle \mu_S^\text{C} \rangle}{ \langle \tilde{\mu}_S^\text{Q} \rangle}
 \ = \  \sqrt{ \frac{S^\text{C}  } {\widetilde{S}^{\text{Q} } } } \times
 		\frac{\sigma_{\tilde{\mu}_S}^\text{Q}} {\sigma_{\mu_S}^\text{C}} \times
			\frac{\langle \mu_S^\text{C} \rangle}{ \langle \tilde{\mu}_S^\text{Q} \rangle}
\label{eq:Ctilde2}
\end{equation}
The relative stability in the mass measurement depends on three important ratios, the relative signal efficiency ($S^\text{C}/\widetilde{S}^{\text{Q} } $), 
the relative fluctuation in the mass spectra 
($\sigma_{\tilde{\mu}_S}^\text{Q}/\sigma_{\mu_S}^\text{C}$), and the relative average mass  
($\langle \mu_S^\text{C} \rangle/ \langle \tilde{\mu}_S^\text{Q} \rangle$). Table~\ref{table:understanding-physics-mass}
exhibits the variation of these quantities with $\alpha$.  As discussed in the previous paragraphs (and indicated in also
Table~\ref{table:understanding-physics-cs}) $\sqrt{S^\text{C}/\widetilde{S}^{\text{Q} }}$ is a slowly but monotonically decreasing
function as $\alpha$ decreases due to the increasing set of tagged jets in the hybrid analysis, \textit{i.e.}, the jets leaking-in at the edge
of the window $\Omega$ as measured by the quantity $\sum_k \Delta_k/\tilde{\epsilon}$ in our simple model.

As shown in Table~\ref{table:understanding-physics-mass}, the average jet mass remains relatively constant ($\langle \tilde{\mu}_S  \rangle \simeq \langle \mu_S^\text{C} \rangle \simeq 80\gev$) for all values of $\alpha$.  In terms of the simple model presented
in the previous section, the shift in the average jet mass (in the window $\Omega$) in going from the conventional
analysis to the hybrid analysis is proportional to the \textit{difference}
between the number of jets leaking-in from the upper edge of the window and the number leaking-in at the lower edge, 
$\left(\sum_{k^+} \Delta_{k^+} - \sum_{k^-} \Delta_{k^-}\right)/\tilde{\epsilon}$ (recall that 
the counting analysis, see Eq.~(\ref{eq:toyNuncer2}), involved the simple sum of these contributions).  Since this
leaking-in process is quite symmetrical (\textit{i.e.}, the signal sample itself is quite symmetrical about $M_W$ with nearly
identical numbers of jets just outside the window at both ends), any shift in
the average jet mass is expected to be quite small, \textit{i.e.}, much smaller than the shift seen in the quantity $\sqrt{S^\text{C}/\widetilde{S}^{\text{Q} }}$, 
in agreement with the results in Table~\ref{table:understanding-physics-mass}.

The last quantity (namely, the the relative fluctuation in the mass spectra, 
$\sigma_{\tilde{\mu}_S}^\text{Q}/\sigma_{\mu_S}^\text{C}$) is especially interesting. 
Table~\ref{table:understanding-physics-mass} shows that this ratio first increases with decreasing $\alpha$, and then
decreases.  The simple model of the previous Section suggests that the size of the deviation from unity for the ratio is
again set by the fraction of tagged jets that are leaking in, $\sum_k \Delta_k/\tilde{\epsilon}$, but now with a coefficient
that, not surprisingly, depends on the \textit{shapes} of the jet mass distributions.   The changes in the mass distribution can
be qualitatively understood as follows.  As $\alpha$ is decreased and we move away from the conventional analysis, the initial change in the mass distribution is
the leaking-in of jets just outside the mass window $\Omega$ into mass bins just inside the window (as is evident in Fig.~\ref{fig:pdfs}).
Thus initially the mass distribution in the hybrid analysis is \textit{broader} than in the conventional analysis and $\sigma_{\tilde{\mu}_S}^\text{Q}/\sigma_{\mu_S}^\text{C}$
increases above unity with decreasing $\alpha$.  However, eventually, as the mass distribution fills in the central region of the window (again see Fig.~\ref{fig:pdfs}),
the Qjets mass distribution again has a width similar to the conventional case and $\sigma_{\tilde{\mu}_S}^\text{Q}/\sigma_{\mu_S}^\text{C}$
goes back to unity (for $\alpha$ just above $0.1$ in Table~\ref{table:understanding-physics-mass}).  With a further decrease of $\alpha$, 
$0.1 > \alpha \geq 0 $, the results in Table~\ref{table:understanding-physics-mass} indicate that the jet mass distribution 
found by the Qjets procedure is \textit{narrower} than the one found by pruning alone, 
{\it i.e.}, the Qjets procedure provides a more efficient groomer than  conventional or classical pruning. 

Overall the relative uncertainty in the tagged mass measurement for the hybrid analysis versus the conventional analysis \textit{decreases} 
with decreasing $\alpha$ and the hybrid result becomes approximately 30\% smaller than 
the fluctuations in the conventional analysis, \textit{i.e.}, it is the $\sqrt{S^\text{C}/\widetilde{S}^{\text{Q} }}$ factor that effectively
controls the $\alpha$ dependence shown in Table~\ref{table:understanding-physics-mass}.  What we have labeled 
the physics part of the mass measurement uncertainty is minimized for $0.01 \geq \alpha \geq 0$.

%------------------------------------------------------------------------------------------------------------------
\section{\label{sec:conclusion}Conclusions}

The Qjets procedure is intuitively motivated by the idea that analyses of jet observables that depend on clustering histories can be improved by considering 
multiple clustering histories of a jet. On the other hand, 
the statistical treatment of the results can be unintuitive and opaque.  Much of the confusion lies in the fact that, while all observables in the Qjets procedure are weighted with weights 
following a continuous distribution in the interval $\left[ 0,1\right]$, the conventional approach applies no weight as long as jets are tagged, \textit{i.e.}, applies a simple binary weight. 
Even in sophisticated multivariate analyses, 
where many variables are combined in a likelihood and each jet/event is assigned a likelihood (a continuous distribution in the interval $\left[ 0,1\right]$) for being a signal, the 
likelihood variable only provides a discriminatory variable  to separate signal from background. The measurements are subsequently estimated from the tagged jet/event sample 
(\textit{i.e.}, the jets/events that pass the cut on likelihood to be signal) with only a binary (0 or 1) weight.  

The purpose of this paper is to address this issue, namely, to provide a platform in which the uncertainties associated with the measurements in the Qjets procedure can be evaluated.  We also propose an alternative way to calculate the uncertainties of measurements. Uncertainties are traditionally estimated using Monte Carlo pseudo-experiments, in which jets/events are picked at random from a given \emph{master-sample} of jets/events (either carefully prepared using a Monte Carlo event generator, or control-samples from collider events), and then repeating pseudo-experiments several times.  Variations of observables over pseudo-experiments then provide an estimate of statistical uncertainties. While this method is 
straightforward, it is time consuming (since pseudo-experiments need to be repeated many times), and still does not provide any insights regarding these measurements. 
In this work we choose a different framework -- we provide analytic formulas in Section~\ref{sec:statuncer}, which relate these uncertainties with various moments of the 
given jets/events sample.  On the one hand, these expressions provide much faster ways to measure uncertainties; while on the other, they help explain the physics of the uncertainties. 
We have also presented a simple model of how the Qjets procedure impacts the probability distributions in both the tagging efficiency $\tau$ and the jet mass $\mu$, which provides
further insight into the observed numerical results.

We find that, while Poisson uncertainties associated with measurements are unavoidable, sampling uncertainties can be reduced by using weighted jets such as those returned by Qjets. We show that this additional stability in measurements provided by the Qjets procedure can arise from two qualitatively different sources -- from the transition from unweighted to weighted measurements (which we label the statistics effects), and from the Qjets generated changes in the distributions of jet-observables themselves, \textit{e.g.}, jet masses, (which we label physics effects). Our explicit numerical results indicate how these two kinds of effects often compete with each other, and how they vary as various Qjets parameters, especially the rigidity $\alpha$, are altered.  Overall, however, the Qjets procedure acts to improve both the statistical stability of counting experiments and the precision of the measurement of jet observables like the jet mass.  Further, we have seen that the Qjets procedure can largely remove the negative impact of a less-than-optimum
choice of jet grooming parameters on a conventional analysis.

Before we conclude, let us note that the results in this work can be  easily generalized. We obtained the expressions for uncertainties only for cross-section and mass measurements. Uncertainties for any other weighted measurements in the Qjets procedure can be performed by following the treatment for the mass measurement. Also note that, in deriving these formulas, we explicitly talked about jets. However, we can easily use the same formalism when we need to talk about events.  In fact, we choose one jet per event in our calculations. Therefore, the expressions for uncertainties associated with the number of jets observed (for example), is identical to the uncertainties associated with the number of events observed. It is straightforward to apply the framework introduced in this work to explain the statistical improvements claimed by the recent proposals such as ``Telescoping Jets"~\cite{Chien:2013kca} and ``Jet Sampling"~\cite{Kahawala:2013sba}.
Finally, we also expect that  sophisticated, state-of-the-art multivariate techniques can be made more robust by 
estimating measurements using weighted events with the likelihood variable as the weight. Such an analysis could presumably follow the framework laid out in this paper.

%------------------------------------------------------------------------------------------------------------------
\section*{Acknowledgements}
The authors would like to thank Matthew D.~Schwartz for his collaboration at an earlier stage of this work. This work was supported in part by the US Department of Energy under contracts DE-FGO2-96ER40956 and DE-SC003916, by a Simons postdoctoral fellowship, Director's fellowships from LANL,  an LHC-TI travel grant and by the KITP, under NSF grant PHY05-51164. The bulk for the computations were performed on the Mapache cluster in the HCP facility at LANL.  Some preliminary computations were also performed on the Odyssey cluster at Harvard University and the TEV cluster in the University of Washington.   

%------------------------------------------------------------------------------------------------------------------
\section*{Note Added}
While this manuscript was being finalized Ref.~\cite{Chien:2014hla} appeared on the arXiv.  Ref.~\cite{Chien:2014hla} studies the statistical effects in counting experiments (\textit{i.e}., cross-sections) for Qjet-like observables using pseudo-experiments in the context of Ref.~\cite{Kahawala:2013sba}. In contrast, in this manuscript we explore both cross-sections and more general measurements such as jet mass and provide an analytical framework for calculating  their statistical properties in terms of probability density functions (the results are then validated using pseudo-experiments in Appendix~\ref{sec:validation}).
%------------------------------------------------------------------------------------------
\appendix
%------------------------------------------------------------------------------------------

\section{\label{sec:validation}Validation of Section~\ref{sec:statuncer} with Pseudo-experiments}

Traditionally,  statistical uncertainties of complicated observables are estimated by using Monte-Carlo pseudo-experiments. 
In this procedure,  one generates many sets of events, where the number of events is chosen according to a Poisson distribution with a given mean (see Eq.~\eqref{eq:pois}). 
One then measures the quantity of interest on each set of events, and, by considering the variation of the quantity across many pseudo-experiments, one can estimate the 
statistical uncertainty of the measurement considered.  This procedure simultaneously accounts for both Poisson and sampling uncertainties.

In this work, we advocate for a different method of calculating statistical uncertainties.  As shown in Section~\ref{sec:statuncer}, analytical expressions may be derived, 
which relate these uncertainties to various moments of a probability distribution constructed from a sample of events. These analytical formulas carry more information 
than just performing Monte-Carlo pseudo-experiments, since they (like all analytical derivations) also explain ``why the numbers are what they are." One can use this improved 
understanding to devise ways to attempt 
to reduce uncertainties further.  

\begin{figure}\centering
	\includegraphics[width=1.0\textwidth]{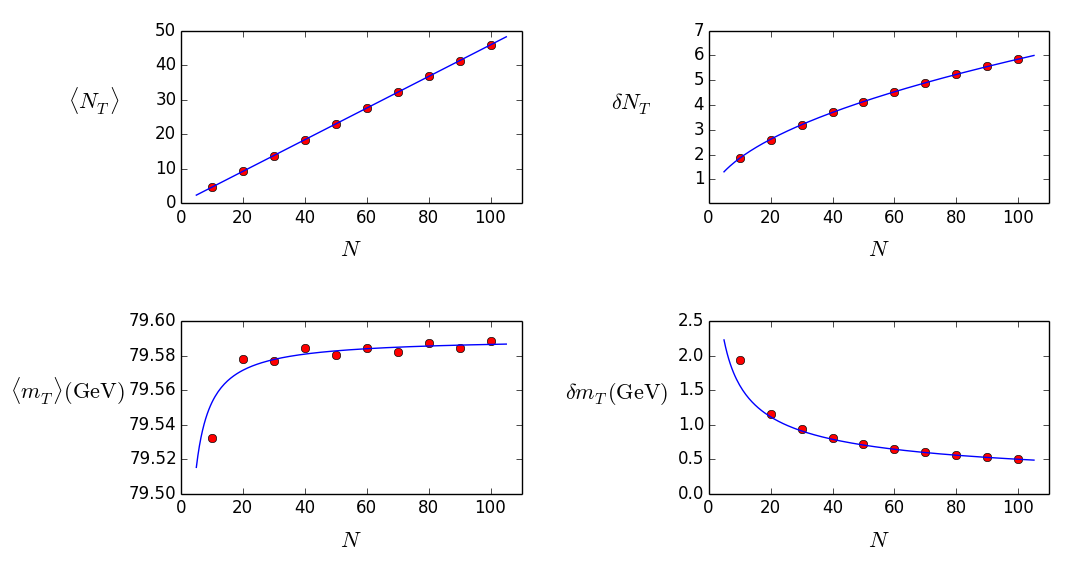}
	\caption{\label{fig:analyticvspexp}  Left: Variation of $\langle N_T \rangle$,  $\delta N_T$, $\langle m_T \rangle$, and  $\delta m_T$ as functions of $N$ for a sample of $W$-jets. 
              See the text for details of the Qjets parameters used. The analytical results (calculated using formulas derived in Section~\ref{sec:statuncer}) are represented by blue lines. 
             The red points denote the same quantities evaluated using Monte Carlo pseudo-experiments.}
\end{figure} 

The purpose of this section of the Appendix is to validate the formulas derived in Section~\ref{sec:statuncer} using pseudo-experiments. In order to do this, we choose a 
sample of $W$-jets (in fact, we choose the same set of hadronic $WW$-events as outlined in Section~\ref{sec:pheno}, and use the same procedure and parameters to 
construct $W$-jets out of these events). We perform $10^5$ pseudo-experiments, in each of which $n$ $W$-jets are chosen at random.  As explained above (see especially  Section~\ref{sec:statuncer}), $n$ follows a Poisson distribution with mean $N$. Jets chosen in a pseudo-experiment, are then subjected to the Qjets procedure for a particular set of parameters ($\alpha= 0.01, z_\text{cut} = 0.1, D_\text{cut} = m/p_T, \Omega = \left( 70-90\right)\gev$). Using the outputs of the Qjets procedure, we calculate observables $N_T$ (number of jets tagged in an experiment) and $m_T$ (tagged mass in the experiment) for each pseudo-experiment. The variations of the these observables over the set of
pseudo-experiments provides estimates of the statistical uncertainties $\delta N_T$ and $\delta m_T$. We also estimate the same uncertainties using the analytic
expressions derived in Section~\ref{sec:statuncer} and compare them. 

In Figure~\ref{fig:analyticvspexp} we compare the analytic estimates of the mean values and the uncertainties, represented by the blue lines, and the numerical values 
from the pseudo-experiments
(the red points) as a function of the mean number of $W$-jets $N$.  We begin with a measurement of the cross section in the top row  of Figure~\ref{fig:analyticvspexp}.  
Here we see that the average and the uncertainty in the number of 
tagged events follow essentially exactly the distribution in Eq.~\eqref{eq:dnnr}.  Next, consider a measurement of the average jet mass, as in the bottom row of  
Figure~\ref{fig:analyticvspexp}.  
Here we see that the uncertainty falls as $1/\sqrt{N}$ up to corrections whose effects are captured by terms of  ${\cal O}(1/N^2)$ in the formulas in Section~\ref{sec:statuncer}.

%------------------------------------------------------------------------------------------
\section{\label{sec:jetmass}Jet Mass}

In this section of the Appendix we derive the analytical expressions relevant for estimating the uncertainties associated with mass. 
In particular, we derive Eqs.~(\ref{eq:mraveNS} - \ref{eq:mrvarNconv}). 
For the sake of completeness, we repeat a few of the definitions introduced in Section.~\ref{subsec:massuncer}.

In an experiment where $N_S$ jets with masses $\{\mu_j \}$ and tagging probabilities $\{\tau_j \}$ are chosen at random, the measured tagged mass is given by
\begin{equation*}
m _T= \frac{ \sum_{ j=1}^{N_S} \mu_j \tau_j}{\sum_{ j=1}^{N_S} \tau_j } \, .
\end{equation*}
We are interested in the average and the variance of $m _T$. Given the fact that the experiment started with $N_S$ jets, we have
\begin{align*}
 \langle m_T \rangle_{N_S} \ & = \  \Big\langle \frac{ \sum_{ j=1}^{N_S} \mu_j \tau_j}{\sum_{ j=1}^{N_S} \tau_j } 
 	 \Big\rangle_{N_S}   \ \equiv \   \Big\langle \frac{M_{T} }{N_T}  \Big\rangle_{N_S}\,,
 		\qquad \qquad \text{ and }   \\
 \left(  \delta m_T \right)^2_{N_S}  \ & = \  
	\langle (m_T - \langle m_T \rangle_{N_S})^2 \rangle_{N_S}  
		\ =  \ \langle m_T^2 \rangle_{N_S} - \langle m_T \rangle_{N_S}^2 \,.
\end{align*}
In these expressions the notation $M_T \equiv m_T N_T$ is used in order to simplify the results.  We note that the probability distribution for $M_T$ and $N_T$,  
for a given sample of size $N_S$, can be constructed in terms of $F_1(\mu, \tau)$,
\begin{equation}
\label{eq:F1mutauNS}
F_{N_S}(M_T,N_T)  =\left[  \prod_{k=1}^{N_S}\int F_1(\mu_k, \tau_k) d\mu_k d\tau_k  \right] \delta\left(N_T - \sum_{k=1}^{N_S} \tau_k\right) \delta\left(M_T - \sum_{k=1}^{N_S} \mu_k \tau_k\right)\,.
\end{equation}
The relevant moments of this general distribution can be derived in terms of the moments of $F_1$
by repeating the manipulations in Eqs.~(\ref{eq:F1NSNT}-\ref{eq:F1NSNT2}).  We have 
\begin{align}
\label{eq:F1mutauNSNT}
\langle N_T \rangle_{N_S}  \ &= \  \int \!  dM_T \, dN_T  \  N_T F_{N_S}(M_T,N_T)  
	 \ = \ N_S \langle \tau \rangle  \\
\label{eq:F1mutauNSMT}
\langle M_T \rangle_{N_S} \ &= \  \int \!  dM_T \, dN_T  \  M_T F_{N_S}(M_T,N_T)   = 
	 	 \ N_S \langle \mu \tau \rangle   \\
 \label{eq:F1mutauNSNT2}
\langle N_T^2 \rangle_{N_S}  \ &= \ \int \!  dM_T \, dN_T  \  N_T^2 F_{N_S}(M_T,N_T)   =  
	N_S^2 \langle \tau \rangle^2 +N_S \sigma_\tau^2 \\
\label{eq:F1mutauNSMT2}
\langle M_T^2 \rangle_{N_S} \ &= \  \int \!  dM_T  \, dN_T  \  M_T^2 F_{N_S}(M_T,N_T)   = 
	N_S^2 \langle \mu \tau \rangle^2 +N_S \sigma_{\mu\tau}^2 \\
\label{eq:F1mutauNSMTNT}
\langle M_T N_T \rangle_{N_S} \ &= \  \int \!  dM_T  \, dN_T  \  M_T N_T F_{N_S}(M_T,N_T)   = 
	N_S^2 \langle \mu \tau \rangle \langle \tau \rangle  + N_S 	 \sigma(\tau, \mu\tau)
\end{align} 

Now we are ready to estimate the mean and variance of  the tagged mass, $m_T$, distribution.  These calculations 
are slightly non-trivial since $m_T$ is a ratio of two independent variables. We use a Taylor series expansion to simplify
the results.  In particular, note that a generic bivariate function $f(x,y)$ can be expanded using
\begin{equation}
\begin{split}
f(x,y)  \simeq &  f(x_0,y_0) +  \frac{\partial f }{\partial x} \Bigg|_{x_0,y_0} \left( x-x_0 \right) 
		+   \frac{\partial f }{\partial y} \Big|_{x_0,y_0} \left(y-y_0 \right)  \\
		& \frac{1}{2} \Bigg[   \frac{\partial^2 f }{\partial x^2} \Bigg|_{x_0,y_0}   \left( x-x_0 \right)^2 \ + \   
			 \frac{\partial^2 f }{\partial y^2} \Bigg|_{x_0,y_0}  \left( y-y_0 \right)^2   
			   \ + \ 2  \frac{\partial^2 f }{\partial x \partial y} \Bigg|_{x_0,y_0}   \left( x-x_0 \right)  \left( y-y_0 \right)  \Bigg] 
			   		 + \dots
\end{split}
\end{equation}
Therefore,  treating $ m_T $ as a function of $M_T$ and $N_T$, we can expand  $ m_T $ around $M_T = \langle M_T \rangle_{N_S}$ and $N_T = \langle N_T \rangle_{N_S}$. We find that 
\begin{equation}
\begin{split}
m_T    \  \simeq \  &
	  \frac{ \langle  M_T \rangle_{N_S}} {\langle N_T \rangle_{N_S}} \ + \  
	  	 \frac{\langle  M_T \rangle_{N_S}} {\langle  N_T  \rangle_{N_S}^3} 
		 	 \left(  N_T - \langle  N_T \rangle_{N_S} \right)^2  \\
	 & \quad - \frac{1}{\langle  N_T  \rangle_{N_S}^2} 
	\left(  N_T - \langle N_T \rangle_{N_S} \right) \left(M_T -  \langle M_T \rangle_{N_S} \right)	
	+ \dots
\end{split}	 
\end{equation}
It is now straightforward to find the average 
\begin{equation}	
 \langle m_T \rangle_{N_S}	\  \simeq  \
	\frac{ \langle \mu \tau\rangle}{\langle \tau \rangle}
		\left[1+ \frac{\sigma_\tau^2}{N_S \langle \tau \rangle^2} 
			- \frac{\sigma(\tau,\mu\tau)}{N_S \langle \mu\tau\rangle \langle \tau\rangle}  \right] + \dots \,.
\end{equation}
A similar expression can be derived for $  m_T^2 $, 
\begin{equation}
\langle m_T^2 \rangle_{N_S}	\  \simeq  \
	\frac{ \langle \mu \tau\rangle^2}{ \langle \tau \rangle^2} \left[1+
		\frac{\sigma_{\mu\tau}^2}{N_S \langle \mu\tau \rangle^2} 
			+ 3 \frac{\sigma_{\tau}^2}{N_S \langle \tau \rangle^2}
				- 4 \frac{\sigma(\tau,\mu\tau)}{N_S \langle \mu\tau\rangle \langle \tau\rangle}  \right]\,.	
\end{equation}
The final step in our calculation involves convolving with the Poisson distributions.  This yields 
\begin{align}
	  \langle m_T \rangle \ &= \ \sum_{N_S=0}^\infty \text{Pois}(N_S|N)  \  \langle m_T \rangle_{N_S}  
	  	\  \simeq \  	\frac{ \langle \mu \tau\rangle}{\langle \tau \rangle} 
		\left[1+ \frac{\sigma_\tau^2}{N \langle \tau \rangle^2}  
			- \frac{\sigma(\tau,\mu\tau)}{N \langle \mu\tau\rangle \langle \tau\rangle}  \right]  \,, \\
	  \langle m_T^2 \rangle \ &= \ \sum_{N_S=0}^\infty \text{Pois}(N_S|N)  \  \langle m_T^2 \rangle_{N_S}  
	  	\  \simeq \   	 	\frac{ \langle \mu \tau\rangle^2}{ \langle \tau \rangle^2} \left[1+
		\frac{\sigma_{\mu\tau}^2}{N \langle \mu\tau \rangle^2} 
			+ 3 \frac{\sigma_{\tau}^2}{N \langle \tau \rangle^2}
				- 4 \frac{\sigma(\tau,\mu\tau)}{N \langle \mu\tau\rangle \langle \tau\rangle}  \right] \,, \\
	\left(  \delta m_T \right)^2 \   &=  \ 
		\langle m_T^2 \rangle - \langle m_T \rangle^2   		  \	  \simeq \
			\frac{ \langle \mu \tau\rangle^2}{N \langle \tau \rangle^2} \left[
		\frac{\sigma_{\mu\tau}^2}{\langle \mu\tau \rangle^2} 
			+  \frac{\sigma_{\tau}^2}{\langle \tau \rangle^2}
				- 2 \frac{\sigma(\tau,\mu\tau)}{\langle \mu\tau\rangle \langle \tau\rangle}  \right]\,.	 		
\end{align}
In these expressions we have neglected terms of order $1/N^2$  and higher. 

Some simplifications arise for the case of the conventional tagging procedure.  Since $\tau$ is non-zero (and equal to one) only in the range $\Omega$, we find that ($q >0$)
\begin{equation}
\left(\langle \mu^p \tau^q \rangle\right)^\text{C} =  \int d\mu \int_0^1 d\tau\, \mu^p \tau^q  F_1^{\text{C}}= 
		 \int_\Omega  d\mu \int_0^1 d\tau\,  \mu^p  F_1^{\text{C}}   
		 	= \left(  N_\Omega \   \langle \mu ^p \rangle \right)^\text{C}
		 		=  \epsilon \  \langle \left(\mu^\text{C} \right)^p \rangle \, ,
\end{equation}
where we use Eq.~\eqref{eq:omegamoments} to derive the final expressions and borrow the notation $\mu^\text{C} $ from Eq.~\eqref{eq:mrvarNconv}, 
to denote that the moment is to be calculated from  Eq.~\eqref{eq:omegamoments} using  the conventional pdf $F_1^\text{C} \left( \mu ,\tau\right)$. 

Therefore we find the following identities  (recall $\langle \tau^\text{C} \rangle = \epsilon$)
\begin{equation}
	\left(\frac{\sigma_{\mu\tau}^2}{\langle \mu \tau \rangle^2}\right)^\text{C} 
=  \frac{\langle \left(\mu^\text{C}\right)^2\rangle  -  \epsilon \langle \mu^\text{C}\rangle^2}
	{\epsilon \langle \mu^\text{C}\rangle^2}
%
%\frac{\left(\sigma_{\mu}^\text{C}\right)^2}{ \langle \mu^\text{C}  \rangle^2},  
\quad \text{and} \quad
	\left(\frac{\sigma(\tau,\mu\tau)}{\langle \mu\tau\rangle \langle \tau\rangle}\right)^\text{C} 
             = \frac{\langle \mu^\text{C} \rangle 
             	(1-\langle \tau^\text{C} \rangle)}{\langle \mu^\text{C} \rangle \langle \tau^\text{C} \rangle}
		= \frac{1-\epsilon}{\epsilon}  
       	=  \frac{ \left( \sigma_{\tau}^\text{C} \right)^2}{\langle \tau^\text{C}  \rangle^2} 	\,.
\end{equation} 
The final expressions for the conventional average mass and its uncertainty then simplify to
\begin{equation}
\begin{split}
	 \langle m_T^\text{C} \rangle \ = \ &  \langle \mu^\text{C}  \rangle  \, , \qquad \qquad   
	 \left( \delta m_T^\text{C}  \right)^2  \ = \ 
	 		\frac{ 1 }{ N  }  \times  \frac{1}{\epsilon} \ 
				\left( \sigma_{\mu}^\text{C} \right)^2  \,,  \\
	 & \left( \frac{\delta m_T^\text{C} } {\langle m_T^\text{C} \rangle} \right)^2 
	 	= \frac{ 1 }{ N  }  \times  \frac{1}{\epsilon} 
			\frac{\left( \sigma_{\mu}^\text{C} \right)^2}{\langle \mu^\text{C}  \rangle^2}  
\end{split}				
\end{equation}

%----------------------------------------------------------------------------------------------------

%\nocite{*}
\bibliographystyle{JHEP}
\bibliography{references}

\end{document}